\documentclass[useAMS,usenatbib]{mn2e}
\usepackage{astrojournals,amsmath,amssymb,graphicx,fleqn,txfonts,color}

\renewcommand*{\d}    {\mathrm{d}}
\newcommand  *{\B}[1] {\boldsymbol{#1}}
\renewcommand*{\H}[1] {\hat{\B{#1}}}
\newcommand*  {\A}[1] {{\langle{#1}\rangle}}
\renewcommand*{\S}[1] {\B{\mathsf{#1}}}
\newcommand*  {\cross}{\times}

\title [Misaligned discs around eccentric binaries]
	   {Misaligned gas discs around eccentric black-hole binaries\\
	   and implications for the final-parsec problem}
\author[Hossam Aly, Walter Dehnen, Chris Nixon, \& Andrew King]
       {Hossam Aly$^1$\thanks{Email: ha183@le.ac.uk,
        walter.dehnen@le.ac.uk}, Walter Dehnen$^1$$^\star$, Chris Nixon$^{2}$\thanks{Einstein Fellow} and Andrew King$^{1, 3}$\vspace{0.05in}\\ 
        $^1$ Department for Physics \& Astronomy, University of Leicester, Leicester LE1 7RH\\
        $^2$ JILA, University of Colorado \& NIST, Boulder CO 80309-0440, USA\\
        $^3$ Astronomical Institute `Anton Pannekoek', University of Amsterdam, Postbus 94249 NL--1090 GE Amsterdam, The Netherlands}
\date{Accepted .
      Received ;
      }
\pagerange{\pageref{firstpage}--\pageref{lastpage}}
\pubyear{2014}
\begin{document}
\maketitle
\label{firstpage}
\begin{abstract}
We investigate the evolution of low mass ($M_{\mathrm{d}}/M_{\mathrm{b}}=0.005$) misaligned gaseous discs around eccentric supermassive black hole (SMBH) binaries. These are expected to form from randomly oriented accretion events onto a SMBH binary formed in a galaxy merger. When expanding the interaction terms between the binary and a circular ring to quadrupole order and averaging over the binary orbit, we expect four non-precessing disc orientations: aligned or counter-aligned with the binary, or polar orbits around the binary eccentricity vector with either sense of rotation. All other orientations precess around either of these, with the polar precession dominating for high eccentricity. These expectations are borne out by smoothed particle hydrodynamics simulations of initially misaligned viscous circumbinary discs, resulting in the formation of polar rings around highly eccentric binaries in contrast to the co-planar discs around circular binaries. Moreover, we observe disc tearing and violent interactions between differentially precessing rings in the disc significantly disrupting the disc structure and causing gas to fall onto the binary with little angular momentum. While accretion from a polar disc may not promote SMBH binary coalescence (solving the `final-parsec problem'), ejection of this infalling low-angular momentum material via gravitational slingshot is a possible mechanism to reduce the binary separation. Moreover, this process acts on dynamical rather than viscous time scales, and so is much faster.
\end{abstract}


%%%%%%%%%%%%%%%%%%%%%%%%%%%%%%%%%%%%%%%%%%%%%%%%%%%%%%%%%%%%%%%%%%%%%%%%%%%%%%%%
\section{Introduction}
It is widely accepted that most massive galaxies host a supermassive black hole (SMBH) at their centre. Galaxy mergers, expected from the hierarchical structure growth scenario based on the $\Lambda$ cold dark matter ($\Lambda$CDM) cosmological model, then result in the formation of SMBH binaries \citep{BegelmanEtal1980}. Both SMBHs of such a binary sink towards the galactic centre due to dynamical friction and form a hard binary \citep{Merritt2001}. However, most such SMBHs appear to be single rather than binary SMBHs, implying that SMBH binaries quickly coalesce and merge. One process driving further binary shrinking are slingshot interactions with stars in the `loss cone': those on orbits intersecting with the binary \citep{SaslawEtal1974}. Since the slingshot mechanism ejects the stars, the `loss cone' needs to be replenished in a relatively short timescale in order to shrink the binary all the way down to separations $\lesssim 10^{-2}$ pc where gravitational waves are expected to drive coalescence. For spherical collisionally relaxed stellar systems, it is thought that the slingshot mechanism stalls well before reaching this separation, resulting in what is known as `the final parsec problem' \citep{MilosavljevicMerritt2003, BerczikEtal2005}. 

Potential stellar dynamical solutions have been sought for gas poor systems. \cite{BerczikEtal2006} studied SMBH binaries evolution in realistic triaxial rotating galaxies and found that the galaxies supply stars on centrophilic orbits refilling the loss cone at a high enough rate to prevent the SMBH binary from stalling and that complete coalescence is achieved in less than 10 Gyr. \cite{KhanEtal2013} found that for axisymmetric galaxies with axis ratio $c/a\leq0.8$ the hardening rate is 25 times faster than for spherical galaxies. Self-consistent \emph{N}-body simulations of merging galaxies containing SMBH found binary hardening rates much higher than idealized spherical models and sufficient to shrink the binary to the gravitational wave coalescence regime \citep{KhanEtal2011, GualandrisMerritt2012}.

Interactions with circumbinary gas discs may change the evolution of the SMBH binary. The mass of such a disc is uncertain and depends on its formation. A galaxy rich merger can channel large amounts of gas towards the centre. If this gas can cool efficiently and avoid fragmentation and substantial star formation, a disc with mass comparable to that of the binary may form.

For a prograde disc, spiral density waves in the disc driven by the outer-Lindblad resonance with the binary transport angular momentum away from the binary \citep{GoldreichTremaine1979}. This mechanism is efficient if the disc reaches very close to the binary, so that it occupies the resonance, and is sufficiently massive for the angular-momentum absorption not to result in an expansion of the inner disc edge
\citep{EscalaEtal2005,MacFadyenMilosavljevic2008}. For a disc with $M_{\mathrm{d}}=0.2M_{\mathrm{b}}$, \cite{CuadraEtal2009} found that binary orbital decay can stall because the disc expands due to absorption of angular momentum from the binary, severely slowing further angular momentum exchange \citep[see also][]{LodatoEtal2009}.

Apart from the classical density-wave mechanism, the infall of gas from the inner edge of the disc into the cavity can be important \citep{RoedigEtal2012,RoedigSesana2014}. The binary may either eject such infalling gas via a gravitational slingshot whereby losing angular momentum and energy, or capture it onto an accretion disc around either component, which adds to the binary angular momentum. The binary evolution is determined by the competition between these two effects and it remains unclear, which one wins in the long term\footnote{This effect was present in the simulations of \cite{CuadraEtal2009} but did not effectuate significant binary evolution. On the other hand, based on an extrapolation to 50 times longer than actually modelled, \cite{RoedigSesana2014} claim efficient binary shrinking. However, since the infall of gas depends on the disc structure at its inner edge, this result is very sensitive to the thermodynamical treatment. \cite{RoedigEtal2012}, for example, found that for isothermal instead of adiabatic gas with an imposed standard $\beta$ cooling prescription, the binary orbital decay can be significantly reduced.}.

For a retrograde co-planar disc, the lack of orbital resonances allows the disc to extend to small radii. This enables the binary to accrete or capture material with negative angular momentum \citep{NixonEtal2011a}. If $M_{\mathrm{d}}\sim M_{\mathrm{b}}$, this may suffice to achieve coalescence \citep{RoedigSesana2014}.

In reality, discs with mass in excess of their aspect ratio times the binary mass are gravitationally unstable and hence, due to the short cooling time in these discs, fragment and form stars much faster than binary coalescence \citep{Gammie2001, Goodman2003, Levin2007}. The numerical treatment of fragmentation, star formation, and stellar feedback is extremely challenging. In all of the aforementioned simulations with such massive discs, these processes have simply been suppressed (by assuming slow cooling which prevents star formation),  overestimating the efficiency of disc-driven binary coalescence. Although star formation will rob the disc of a significant amount of gas, the newly formed stars may still contribute to binary orbital decay
\citep[e.g.][]{SesanaEtal2007, SesanaEtal2008}, though less so than the gas owing to the lack of an efficient dissipation mechanism to reduce their pericentres.

A more likely scenario than binary coalescence driven by the interaction with a single massive disc is the repeated interaction with low-mass discs resulting from the infall
and tidal disruption of molecular clouds onto the binary. Such discs are expected to have masses $10^{5-6}\,$M$_\odot$ typical of molecular clouds, small compared to the typical mass $10^{6-9}\,$M$_\odot$ of a SMBH binary. \cite{NixonEtal2011a} studied retrograde discs of this type and found that they are very efficient in reducing the binary angular momentum through accretion of gas with negative angular momentum onto the secondary black hole. This enhanced accretion onto the secondary black hole increases the binary's eccentricity, decreasing the pericentre distance in the process, and coalescence is achieved when a mass comparable to the secondary black hole has been accreted. 

If accretion events in galactic nuclei are chaotic and randomly oriented \citep{KingPringle2006, KingPringle2007, KingEtal2008}, we expect the formation of misaligned circumbinary discs around SMBH binaries. In the case of a circular SMBH binary, the interaction between the misaligned disc and the binary is similar to Lense-Thirring precession on an accretion disc around a spinning black hole \citep{BardeenPetterson1975, Pringle1992, ScheuerFeiler1996}. \cite{KingEtal2005} showed that the induced differential precession will cause a misaligned disc to counter-align with the black hole spin provided
\begin{equation}
    \cos\theta < \frac{-|\B{J}_{\mathrm{d}}|}{2|\B{J}_{\mathrm{h}}|}
\end{equation}
where $\B{J}_{\mathrm{d}}$ and $\B{J}_{\mathrm{h}}$ are the disc and black hole angular momenta, respectively; and $\theta$ is the angle between them. The disc will co-align with the black hole spin if this relation is not satisfied. \cite{NixonEtal2011b} showed that the same analysis applies to the case of a misaligned disc around a binary
(though the precession rate is slightly different). Thus, if counter-alignment is stable \citep{Nixon2012}, this mechanism can provide a solution to the final parsec problem by supplying retrograde discs to achieve coalescence.

Recently, \cite{NixonEtal2013} performed 3-D hydrodynamical simulations of circumbinary discs around a circular binary for various tilt angles $\theta$. In addition to co- and counter-alignment, they found that in many cases the discs is torn into distinct rings which precess almost independently \citep{NixonEtal2012}. The precessing rings, which have partially opposed angular momentum, may interact causing partial cancellation of their angular momenta and thus gas infall close to the binary. This disc tearing significantly increases the accretion rate and may play an important role in promoting the binary final coalescence.

Those studies considered the case of a circular binary interacting with a circumbinary disc, when disc precession is only around the pole of the binary plane. In this study, we consider the more general situation of an eccentric binary. For a SMBH binary formed via a galaxy merger, we expect high eccentricities in many cases \citep{Aarseth2003, KhanEtal2011, WangEtal2014}. Moreover, retrograde accretion onto a circular binary naturally results in eccentricity growth as discussed earlier. One important effect of binary eccentricity is to make the time averaged binary potential triaxial rather than axisymmetric as for a circular binary. Previous studies have shown that misaligned discs in triaxial galaxies can precess around both the major and the minor axes \citep{CameronDurisen1984, ThomasEtal1994}. 

The paper is organised as follows. In Section~\ref{sec:quadrupole} we present analytic results for a simple orbit-averaged model for the binary-disc interaction up to quadrupole order. Section~\ref{sec:simulations} describes the setup of our 3D hydrodynamical simulations, the of which results are presented in Section~\ref{sec:results} and discussed in Section~\ref{sec:discuss}. Finally, we summarise and conclude in Section~\ref{sec:conclusion}.

%%%%%%%%%%%%%%%%%%%%%%%%%%%%%%%%%%%%%%%%%%%%%%%%%%%%%%%%%%%%%%%%%%%%%%%%%%%%%%%%
\section{Binary-Disc quadrupole interaction}
\label{sec:quadrupole}
The dynamics of a circumbinary gaseous ring orbiting an eccentric binary is not analytically treatable, even without considering any dissipation. However, useful insight can be obtained by (i) truncating the binary gravitational potential at quadrupole order, (ii) assuming that the ring is circular, (iii) time-averaging over the binary orbit, and (iv) neglecting dissipation. Assumptions (i) and (ii) are valid as long as the ring is sufficiently distant from the binary, while assumption (iii) requires that orbital resonances between ring and binary are not important.

The monopole of the gravitational interaction results in Keplerian motion of the ring around the binary centre of mass, while the quadrupole describes the lowest-order deviation of the binary from a central point mass.

Recently, \cite{NaozEtal2013} have used Hamiltonian perturbation theory to obtain the equations for the secular evolution of a hierarchical triple up to octopole order.
For a circular outer binary, their results are equivalent to the situation of a circular circum-binary ring. We now summarise the relevant relations (obtained in Appendix~\ref{app:quadrupole} with Newtonian dynamics, but otherwise equivalent to those of
\citeauthor{NaozEtal2013}) in terms of vectors rather than orbital elements to describe the system.

The binary is parametrised by its mass ratio $q\equiv m_2/m_1\le1$, total mass $M=m_1+m_2$, semi-major axis $a$, specific angular momentum $\B{h}$, and eccentricity vector $\B{e}$. Let $\B{R}\equiv\B{x}_1-\B{x}_2$ the instantaneous binary separation vector, then
\begin{eqnarray}
	\B{h} &=& \frac{m_1}{M}\B{x}_1\cross\dot{\B{x}}_1
	+ \frac{m_2}{M}\B{x}_2\cross\dot{\B{x}}_2
	= \frac{q}{(1+q)^2}\B{R}\cross\dot{\B{R}}
	\quad\text{and}
	\\
	\B{e} &=& \frac{\dot{\B{R}}\cross(\B{R}\cross\dot{\B{R}})}{GM} - \hat{\B{R}}.
\end{eqnarray}
The vector $\B{e}$ is conserved for the binary orbit and points from the centre of mass to peri-apse (hence is always orthogonal to $\B{h}$). Its magnitude is the
orbital eccentricity and is related to that of $\B{h}$ by
\begin{equation} \label{eq:h:e}
	h^2(1+q)^4=q^2GMa(1-e^2).
\end{equation}

%%%%%%%%%%%%%%%%%%%%%%%%%%%%%%%%%%%%%%%%%%%%%%%%%%%%%%%%%%%%%%%%%%%%%%%%%%%%%%%%
\subsection{Ring evolution}
The circular circumbinary ring is parametrised by its mass $m$, radius $r$, and pole $\H{l}$. The latter is the unit vector in direction of the ring's specific angular momentum $\B{l}$, which has amplitude $l=\sqrt{G(M+m)r}$. The ring radius must satisfy $r>a(1+e)/(1+q)$ for the quadrupole-approximation to be valid (and in order to avoid collision with a binary component). Note that the tilt angle $\theta$ of the ring with respect to the binary satisfies $\cos\theta=\H{l}{\cdot}\H{h}$.

The quadrupole interaction energy between binary and ring, averaged over both the binary orbit and the ring, is
\begin{equation}
	\label{eq:E}
	\A{E_{\mathrm{br}}} = 
	-\frac{m\omega^2a^2q}{8(1+q)^2} \left[6e^2-1
	- 15e^2(\H{l}\cdot\H{e})^2 + 3(1-e^2)(\H{l}\cdot\H{h})^2\right]
\end{equation}
with $\omega=\sqrt{G(M+m)/r^3}$ the orbital frequency of the ring
, in agreement with equation~(22) of \cite{NaozEtal2013}. The time-averaged binary quadrupole torques the ring according to
\begin{equation} \label{eq:torque:l}
	\dot{\B{l}} = \B{\Theta} \cross \B{l}
\end{equation}
with the vector
\begin{equation} \label{eq:Theta}
	\B{\Theta}
	= \frac{3\omega q}{4(1+q)^2}\frac{a^2}{r^2}
	\left[
	5e^2(\H{l}\cdot\H{e})\,\H{e}
	-(1-e^2)(\H{l}\cdot\H{h})\,\H{h}
	\right].
\end{equation}
From equation~(\ref{eq:torque:l}) we have $\B{l}\cdot\dot{\B{l}}=0$, i.e.\ $\dot{l}=0=\dot{r}$ and the ring is merely precessing \cite[this is no longer true at octopole and higher order, when $\dot{l}\neq0$, see][]{NaozEtal2013}.

Since $\omega m r^2\B{\Theta}=\partial\A{E_{\mathrm{br}}}/\partial\H{l}$, equation~(\ref{eq:torque:l}) implies
\begin{equation} \label{eq:dot:l.dE}
	\frac{\d\H{l}}{\d t}\cdot\frac{\partial\A{E_{\mathrm{br}}}}{\partial\H{l}}=0.
\end{equation}
Thus, (in the assumed approximation) no energy is exchanged between binary and a single ring, but only angular momentum (if the binary interacts with several rings, the individual interaction energies with each ring are no longer conserved).

%%%%%%%%%%%%%%%%%%%%%%%%%%%%%%%%%%%%%%%%%%%%%%%%%%%%%%%%%%%%%%%%%%%%%%%%%%%%%%%%
\subsection{Binary evolution}
The torque of the binary from the ring can be worked out analogously to that of the ring from the binary. After averaging over the binary orbit, we obtain
\begin{equation} \label{eq:H:dot}
	\dot{\B{h}} = - \frac{m}{M} \B{\Theta} \cross \B{l}.
\end{equation}
In particular, the total angular momentum, $M\B{h}+m\B{l}$, is conserved at quadrupole order. For the case $m\ll M$ considered here, the orientation $\H{h}$ only varies slightly even if the disc orientation $\H{l}$ undergoes large changes.

For a circular binary $\B{\Theta}$ is parallel to $\H{h}$ such that $\dot{\B{h}}{\cdot}\B{h}=0$, i.e. $h=|\B{h}|$ is conserved and the binary is merely precessing (with an amplitude that is smaller than that of the disc by a factor $m/M$). This fact together with conservation of total angular momentum was the basis of the analysis by \cite{NixonEtal2011b}.

For an eccentric binary, the evolution of $\B{h}$ is not simply a precession and $h$ not conserved. Instead, we find 
\begin{equation} \label{eq:h:dot}
	\dot{h} = -\frac{15}{4}  \frac{\omega^2 m}{\Omega M}
	\frac{e^2h}{\sqrt{1-e^2}}
	\,(\H{l}\cdot\H{e})\;(\H{l}\cdot\H{k})
\end{equation}
with $\Omega=\sqrt{GM/a^3}$ the binary orbital frequency. Thus, $h$ remains unchanged only if $e=0$ (circular binary), or if $\H{l}$ is perpendicular to either $\H{e}$ or $\H{k}$, i.e.\ if either $\H{e}$ or $\H{k}$ are in the ring plane. Otherwise, $h$ oscillates, since $\H{l}\cdot\H{k}$ oscillates around zero under the ring precession.

The change of the eccentricity vector is
\begin{eqnarray}
	\dot{\B{e}}
	&=& \frac{3}{4}\frac{\omega^2 m}{\Omega M}e\sqrt{1-e^2} \bigg\{
	\left[2-(\H{l}\cdot\H{h})^2-5(\H{l}\cdot\H{e})^2\right]\,
	\H{k}
	+\nonumber \\ &&
	\phantom{3\frac{\omega^2 m}{\Omega M}e\sqrt{1-e^2} \bigg\{}
	+(\H{l}\cdot\H{k})(\H{l}\cdot\H{h})\,\H{h}
	\label{eq:evec:dot}
	%\phantom{3\frac{\omega^2 m}{\Omega M}e\sqrt{1-e^2} \bigg\{}
	+5(\H{l}\cdot\H{e})(\H{l}\cdot\H{k})\,\H{e}
	\bigg\}
\end{eqnarray}
and the corresponding change in eccentricity
\begin{equation} \label{eq:e:dot}
	\dot{e} = \frac{15}{4} \frac{\omega^2 m}{\Omega M}e\sqrt{1-e^2}
	(\H{l}\cdot\H{e})(\H{l}\cdot\H{k}).
\end{equation}
in agreement with equation~A34 of \citealt{NaozEtal2013}, but also with 
equation~(\ref{eq:h:dot}) in conjunction with equation~(\ref{eq:h:e}). In addition to the precession of the orbital plane and the oscillation of the eccentricity (both already described by equation~\ref{eq:h:dot}), the binary also undergoes apsidal precession with rate
\begin{equation} \label{eq:pas:prec}
	\dot{\H{e}}\cdot\H{k} =  \frac{3}{4}\frac{\omega^2 m}{\Omega M}\sqrt{1-e^2}
	\left[2-(\H{l}\cdot\H{h})^2-5(\H{l}\cdot\H{e})^2\right],
\end{equation}
which is prograde for near-planar disc orientations (when $|\H{l}\cdot\H{h}|\sim1$), but retrograde for near-polar discs (when $|\H{l}\cdot\H{e}|\sim1$).

%%%%%%%%%%%%%%%%%%%%%%%%%%%%%%%%%%%%%%%%%%%%%%%%%%%%%%%%%%%%%%%%%%%%%%%%%%%%%%%%
\subsection{Ring precession}

The rates of change of the directions of the binary and ring angular momenta satisfy
\begin{equation} \label{eq:dhdt:dldt}
	\left|\frac{\d\H{h}}{\d t}\right| \le
	\frac{ml}{Mh} \left|\frac{\d\H{l}}{\d t}\right|
\end{equation} 
and a similar relation holds for $|\d\H{e}/\d t|$. Thus, as long as $m\ll M$, the binary orientation changes only very little and/or much more slowly than that of the
ring (except for extreme binary eccentricities when $Mh\propto\sqrt{1-e^2}$ can be small). We therefore consider in this subsection the limit $m/M\to0$ when the binary orientation and eccentricity are constant.

Then, equation~(\ref{eq:dot:l.dE}) implies that the ring precesses along curves of constant $\A{E_{\mathrm{br}}}$. Isolated minima and maxima of $\A{E_{\mathrm{br}}}$ denote stable, non-precessing ring orientations. In the presence of dissipation (due to viscosity in the disc), these orientations are attractors, i.e.\ the dissipative damping  of the precession eventually aligns the pole $\H{l}$ of the ring with the extrema of $\A{E_{\mathrm{br}}}$ \citep*{CameronDurisen1984}. For any $e<1$, the orientations $\H{l}=\pm\H{h}$ are isolated minima of $\A{E_{\mathrm{br}}}$ and correspond to co-planar ring orientations either co- or counter-rotating with the binary. 
%
% Fig 1 double column
%
\begin{figure*}
  \begin{center}
    \resizebox{42.0mm}{!}{\includegraphics{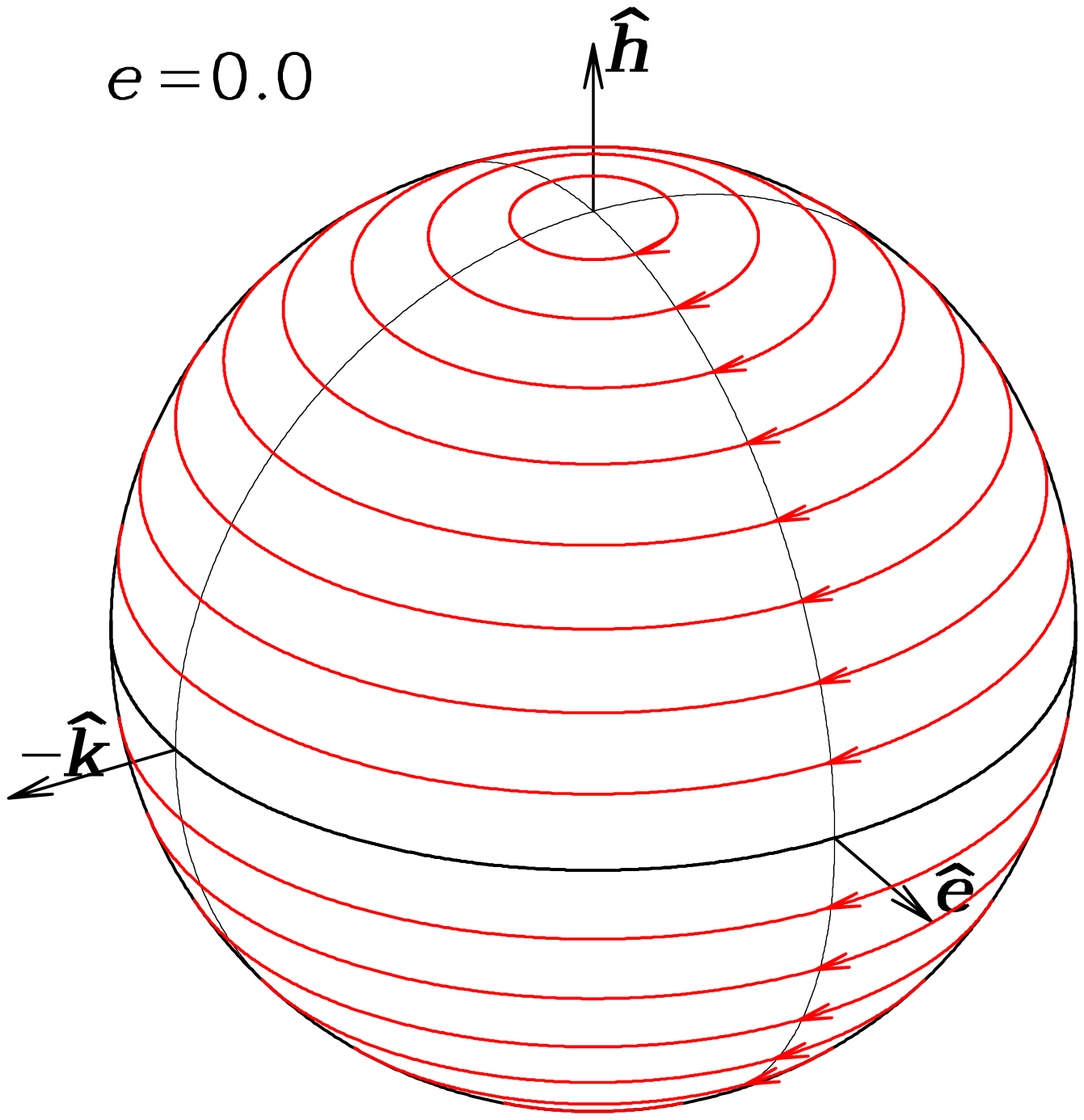}}\hfil
    \resizebox{42.0mm}{!}{\includegraphics{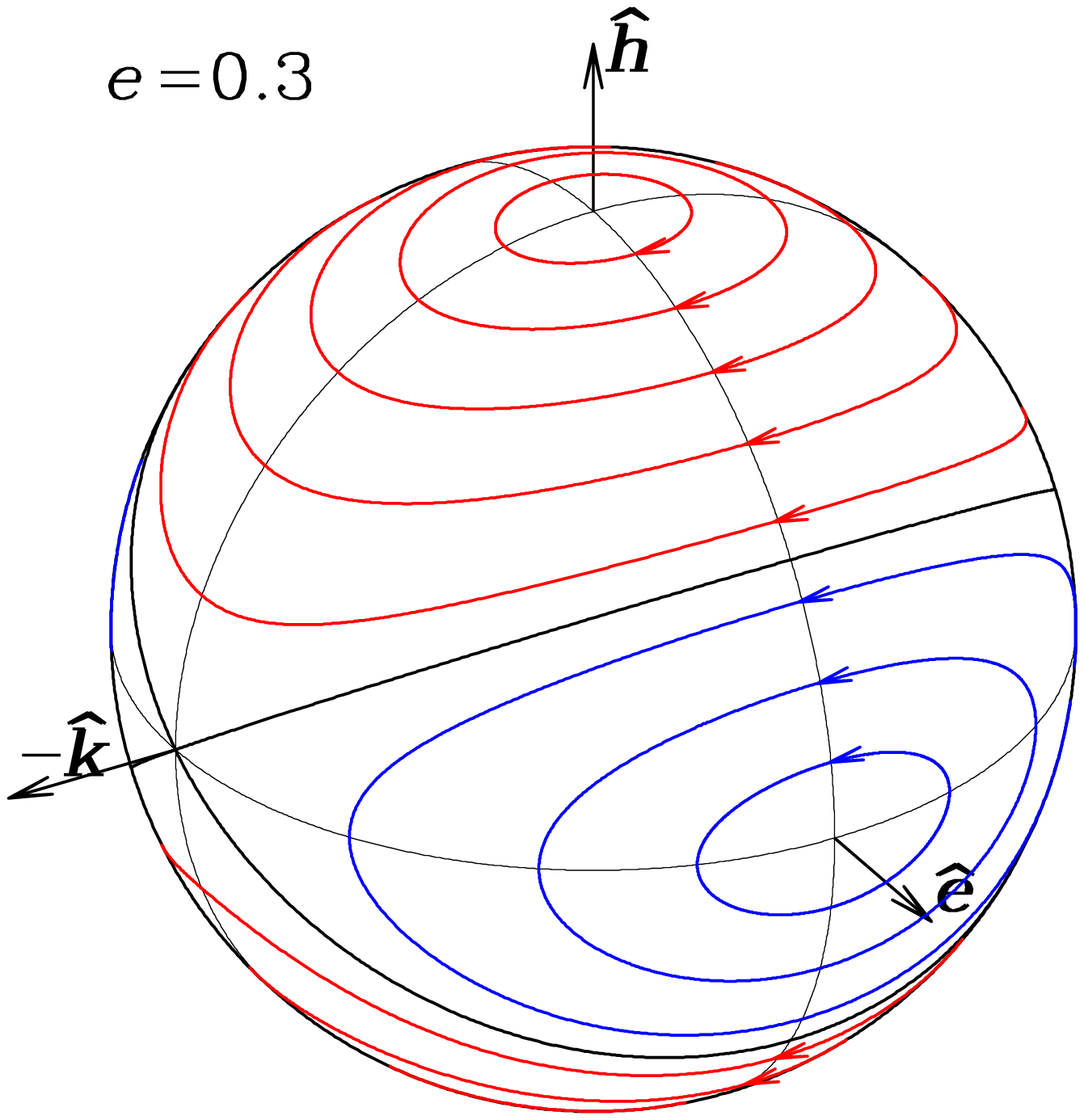}}\hfil
    \resizebox{42.0mm}{!}{\includegraphics{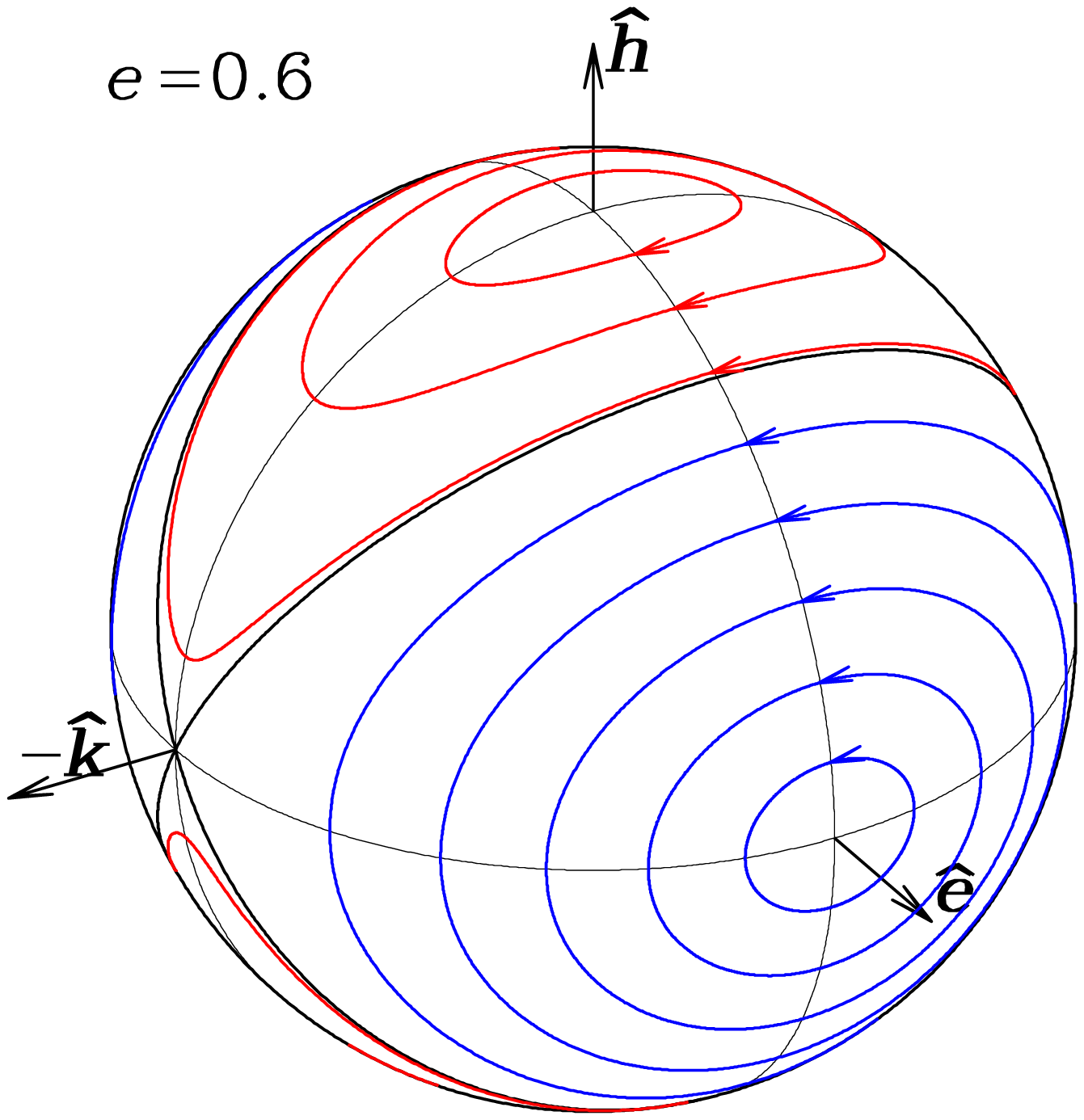}}\hfil
    \resizebox{42.0mm}{!}{\includegraphics{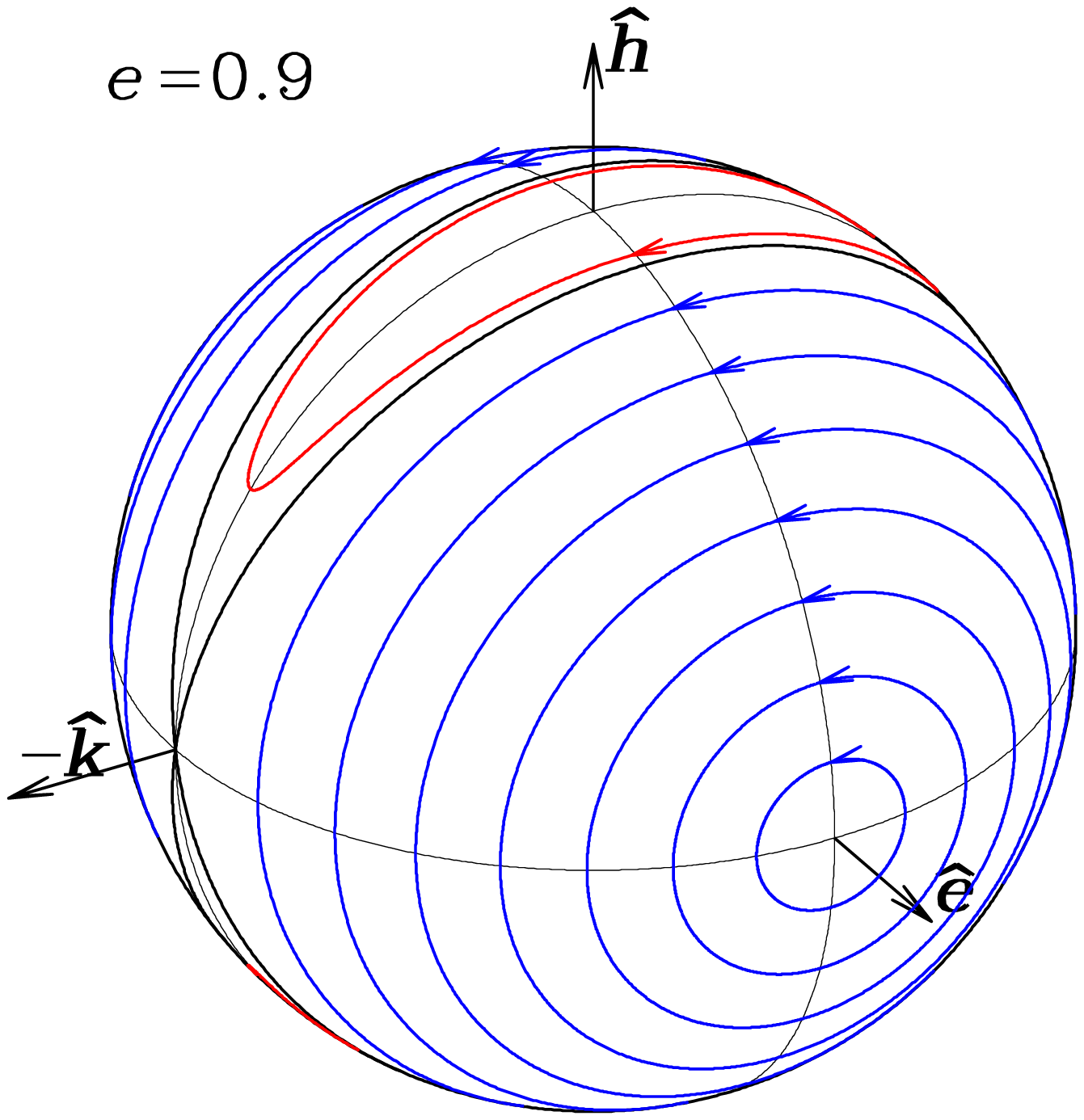}}
    \caption{\label{fig:prec}
    Precession paths for the direction $\H{l}$ of the angular momentum of a dissipation-less circular ring of negligible mass orbiting a binary with eccentricity $e$ as indicated. The binary orbits counter clockwise in the plane perpendicular to its specific angular momentum $\B{h}$ with peri-apse in the direction $\H{e}$. For a circular binary ($e=0$, left), $\H{l}$ always precesses around $\B{h}$ in a retrograde sense. For eccentric binaries, prograde polar precession (blue) around $\B{e}$, the long axis of the time-averaged binary potential, is also possible. The regions of polar and azimuthal precession are separated by two great circles (black). The four ring orientations $\H{l}=\pm\H{h}$ and $\H{l}=\pm\H{e}$ are stable (non-precessing), while the orientations $\H{l}=\pm\H{k}$ are unstable. Dissipation would damp the precession and eventually align the ring with one of the four stable orientations. In case of a massive ring, the binary orbit evolves too: the vectors $\B{h}$ and $\B{e}$ oscillate and precess, and $\B{e}$ and $\H{k}$ rotate around $\B{h}$.}
  \end{center}
\end{figure*}
%%%

For a circular binary ($e=0$), these are the only stable orientations, but all polar orbits ($\theta=90^\circ$) maximise $\A{E_{\mathrm{br}}}$. $\B{\Theta}$ is parallel to $\B{h}$ and ring precession is circular: $\H{l}$ describes a circle around either of the stable orientations, see also the left plot in Fig.~\ref{fig:prec}. The precession rate is lower than the orbital frequency by the factor $3qa^2\cos\theta/4r^2(1+q)^2$. This is the situation previously studied by \cite{NixonEtal2011b}. We now turn to the more general case of an eccentric binary.

For $e>0$, the orientations $\H{l}=\pm\H{e}$ are maxima of $\A{E_{\mathrm{br}}}$, corresponding to polar rings (with opposite senses of rotation) around $\H{e}$. For
$e<1$, $\H{l}=\pm\H{k}$ are saddle points and correspond to polar rings around \begin{equation} \label{eq:k}
	\H{k}\equiv \H{h}\cross\H{e},
\end{equation}

the intermediate axis of the time-averaged binary potential. These latter ring orientations are unstable, i.e.\ small deviations will result in precession around either of the four stable orientations. For $0<e<1$, ring precession is never circular: $\H{l}$ describes a curve elongated towards the unstable orientations, rather than a circle. Azimuthal and polar precessions are retrograde and prograde, respectively.
See Fig.\ref{fig:prec} for a visualisation of the precession paths.

\begin{figure}
  \begin{center}
  	\resizebox{80mm}{!}{\includegraphics[angle=-90]{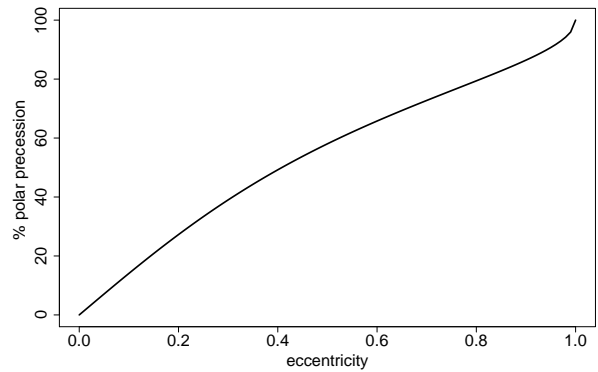}}
	\caption{\label{fig:fraction_polar}Percentage of ring orientations undergoing polar precession as a function of binary eccentricity.}
  \end{center}
\end{figure}

The regions of polar and azimuthal precession are separated by the contours of $\A{E_{\mathrm{br}}}$ passing through the saddle points. These separatrices are circular 
and shown as black in Fig.~\ref{fig:prec}. The fraction of ring orientations undergoing polar precession is
\begin{equation}
	\frac{1}{\pi}\,\cos^{-1}\frac{1-6e^2}{1+4e^2}.
\end{equation}
At small $e$, this grows linearly ($\propto\sqrt{20}e/\pi$) with eccentricity (see Fig.~\ref{fig:fraction_polar}). Azimuthal and polar precession are equally likely for $e=6^{-1/2} \approx 0.408$.

%
% Fig. 2
%
%%%

%%%%%%%%%%%%%%%%%%%%%%%%%%%%%%%%%%%%%%%%%%%%%%%%%%%%%%%%%%%%%%%%%%%%%%%%%%%%%%%%
\section{Simulation setup}
\label{sec:simulations}
We perform a set of 3-D Smoothed Particle Hydrodynamics (SPH) \citep{GingoldMonaghan1977, Lucy1977} simulations of geometrically thin accretion discs with different initial misalignment around an eccentric binary. We use a range of different binary eccentricities $e=0$, 0.3, 0.6, and 0.9. The disc setup is very similar to that used by \cite{NixonEtal2013}: the disc is initially flat and extends from an inner radius of $2a$ to an outer radius of $8a$ with an inner thickness $H/R=0.01$. We use a disc viscosity coefficient \citep{ShakuraSunyaev1973} $\alpha=0.1$ which we setup using an appropriate SPH artificial viscosity coefficient $\alpha_{AV}$ corresponding to our resolution \citep{LodatoPrice2010}. All simulations start with 4 million SPH particles, while the the binary is modelled using two equal mass sink particles with accretion radius of $0.05a$. The disc initial surface density follows the profile $\Sigma \propto R^{-3/2}$, and we use a locally isothermal equation of state with sound speed $c_s\propto R^{-3/4}$. These choices ensure a uniform vertical resolution \citep[and hence uniform physical viscosity, see][]{LodatoPringle2007}. We assume a
disc mass of $M_{\mathrm{d}}/M_{\mathrm{b}}=0.005<H/R$, which ensures that disc self-gravity is not important (we do not include gas self-gravity in our simulations, but we do self-consistently include the back-reaction from the gas on the binary). The simulations were performed using our own code \citep{DehnenAly2012}, which implements an SPH scheme very similar to that used by \cite{NixonEtal2013}, and we verified that our results for $e=0$ agree with theirs.
%
% Fig.3 double column, to be on the same page as the begin of section 4
%
\begin{figure*}
\begin{minipage}{\textwidth}
	\hspace{23.6 mm} $\theta=30^\circ$
	\hspace{24.2 mm} $\theta=45^\circ$
	\hspace{24.2 mm} $\theta=60^\circ$
	\hspace{24.2 mm} $\theta=80^\circ$
	\hspace{24.2 mm} $\theta=90^\circ$ \vspace{2 mm}
\end{minipage}
\begin{minipage}[c]{0.06\textwidth}
	\vspace{15 mm}	  $e=0$
	\vspace{30 mm} \\ $e=0.3$
	\vspace{30 mm} \\ $e=0.6$
	\vspace{30 mm} \\ $e=0.9$ \\ \vspace{15 mm}
\end{minipage}%
\begin{minipage}{0.94\textwidth}
	\includegraphics[angle=0,width=\textwidth]{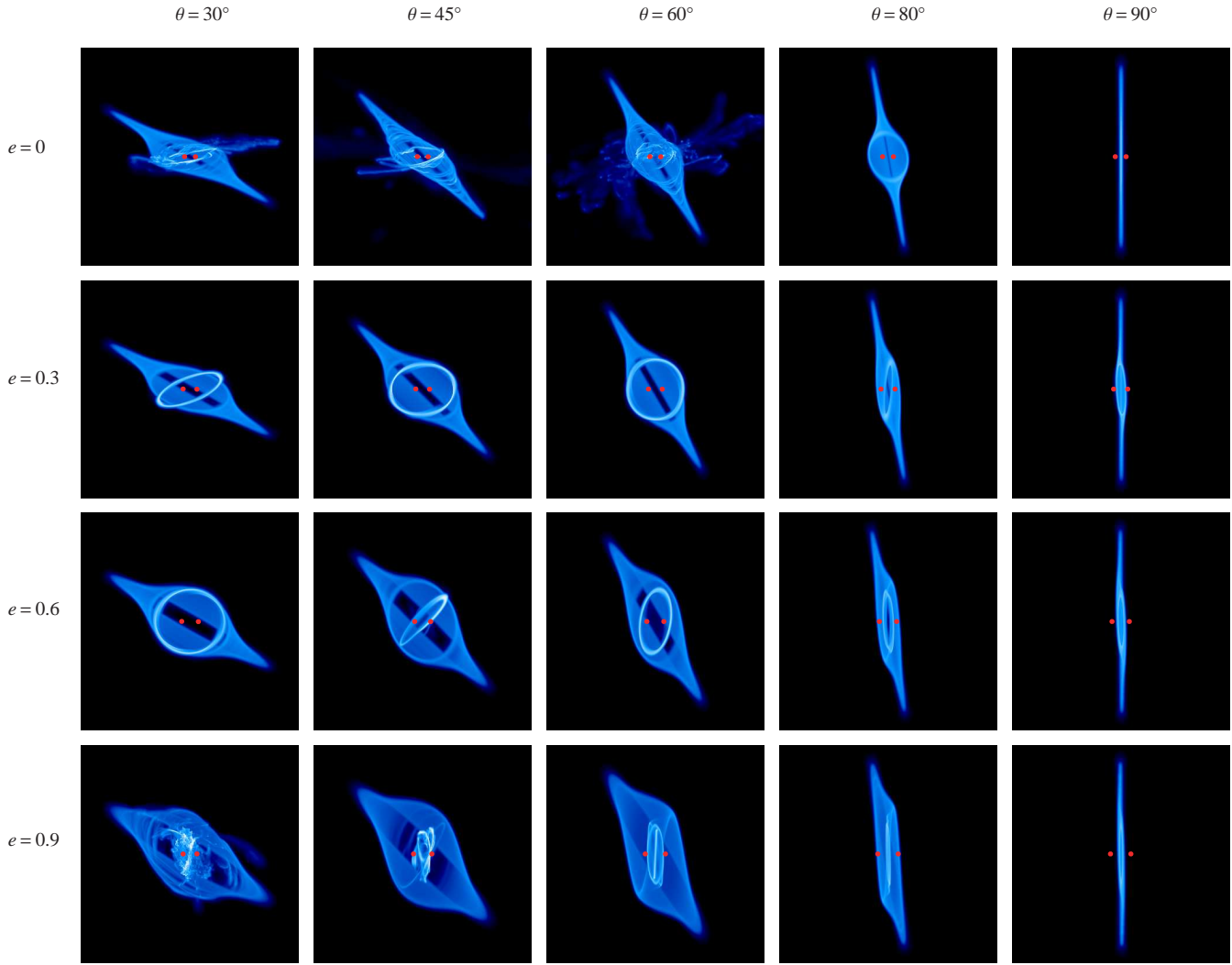}
\end{minipage}
	\caption{\label{fig:render}
	Density rendering of the simulations after $t=600$ ($\approx 95$ binary orbits)
	for different binary eccentricities and initial misalignment angles as indicated.
	The projections are along the intermediate binary axis $\B{k}$ with the angular-		momentum vector $\B{h}$ pointing upwards and the eccentricity vector $\B{e}$ to the 	right.}
\end{figure*}
%%%
The disc has initial angular momentum direction
\begin{equation}
	\H{l}=\sin\theta\cos\phi\,\H{e}+\sin\theta\sin\phi\,\H{k}+\cos\theta\,\H{h},
\end{equation}
where $\phi$ is the \emph{twist} angle of the disc. 
We ran a total of 118 simulations for $e=0$, 0.3, 0.6, and 0.9;
$\theta=0^\circ$, $10^\circ$, $30^\circ$, $45^\circ$, $60^\circ$, $80^\circ$, $90^\circ$, $100^\circ$, $120^\circ$, $135^\circ$, $150^\circ$, $170^\circ$, and $180^\circ$;
$\phi=0^\circ$, $-45^\circ$, and $-90.^\circ$. In the next section  $\phi$ will be taken to be zero whenever it is not specified, we discuss the effects of varying $\phi$ separately.

We point out that the choice of a locally isothermal equation of state implies that the disc instantly radiates away all the heat gained from the viscous dissiaption and shocks. This is justified if the the cooling time is much shorter than the precession time. When this assumption does not hold, the disc thickness will increase and will be more able to resist breaking. We leave more advanced thermodynamic treatment to future investigation.

%%%%%%%%%%%%%%%%%%%%%%%%%%%%%%%%%%%%%%%%%%%%%%%%%%%%%%%%%%%%%%%%%%%%%%%%%%%%%%%%
\section{Simulation Results}
\label{sec:results}
Fig.~\ref{fig:render} shows snapshots after $\approx 95$ binary orbits for the twenty simulations with initial tilt angles $\theta=30^\circ$, 45$^\circ$, 60$^\circ$, 80$^\circ$, and 90$^\circ$ initial binary eccentricities $e=0$, 0.3, 0.6, and 0.9. As expected from the analysis in Section~\ref{sec:quadrupole}, the disc precesses around $\H{h}$ for circular and low-eccentricity binaries, while for very high eccentricities the precession is predominantly around $\H{e}$. In almost all cases, the disc breaks into distinct rings, which interact with each other and, depending on the details of each case, result in either co-, counter-, or polar-alignment of the disc. In some cases the interaction between independently precessing such rings is very violent and disruptive, leading to ejection of gas. We now visit each possible outcome in detail.

%%%%%%%%%%%%%%%%%%%%%%%%%%%%%%%%%%%%%%%%%%%%%%%%%%%%%%%%%%%%%%%%%%%%%%%%%%%%%%%%
\subsection{Polar alignment}
\cite{NixonEtal2011b} showed that for a circular binary, where the binary induced precession is only around $\H{h}$, the disc eventually co- or counter-aligns with the binary orbital plane depending on the disc angular momentum and its initial misalignment angle. Our analysis in Section~\ref{sec:quadrupole} suggests that for an eccentric binary the precession will be around $\H{h}$ or $\H{e}$. In the latter case, dissipation results in polar-alignment.
%
% Fig. 4
%
\begin{figure}
  \begin{center}
    \resizebox{81.0mm}{!}{\includegraphics[angle=-90]{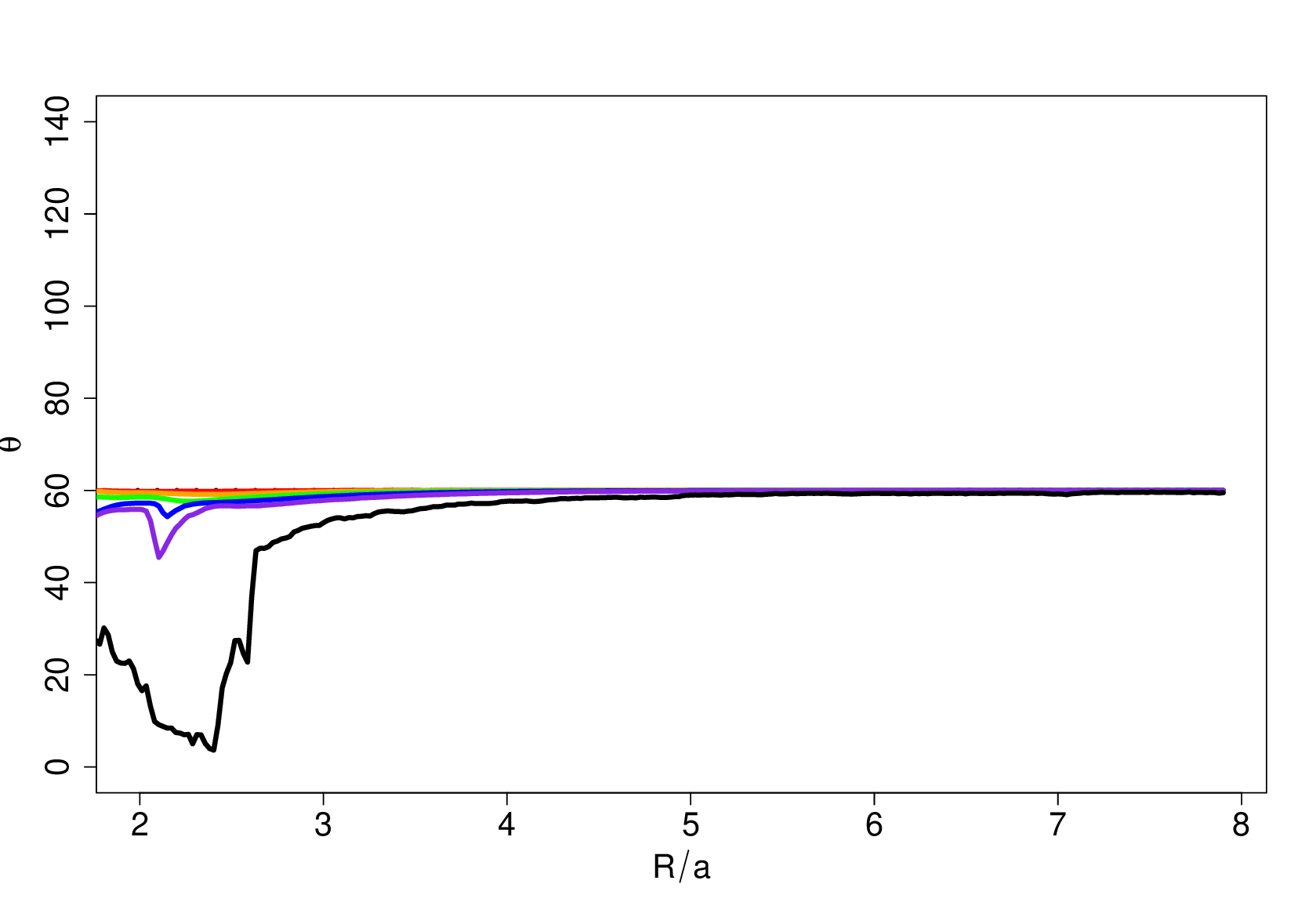}}
	\\[-4ex]
    \resizebox{81.0mm}{!}{\includegraphics[angle=-90]{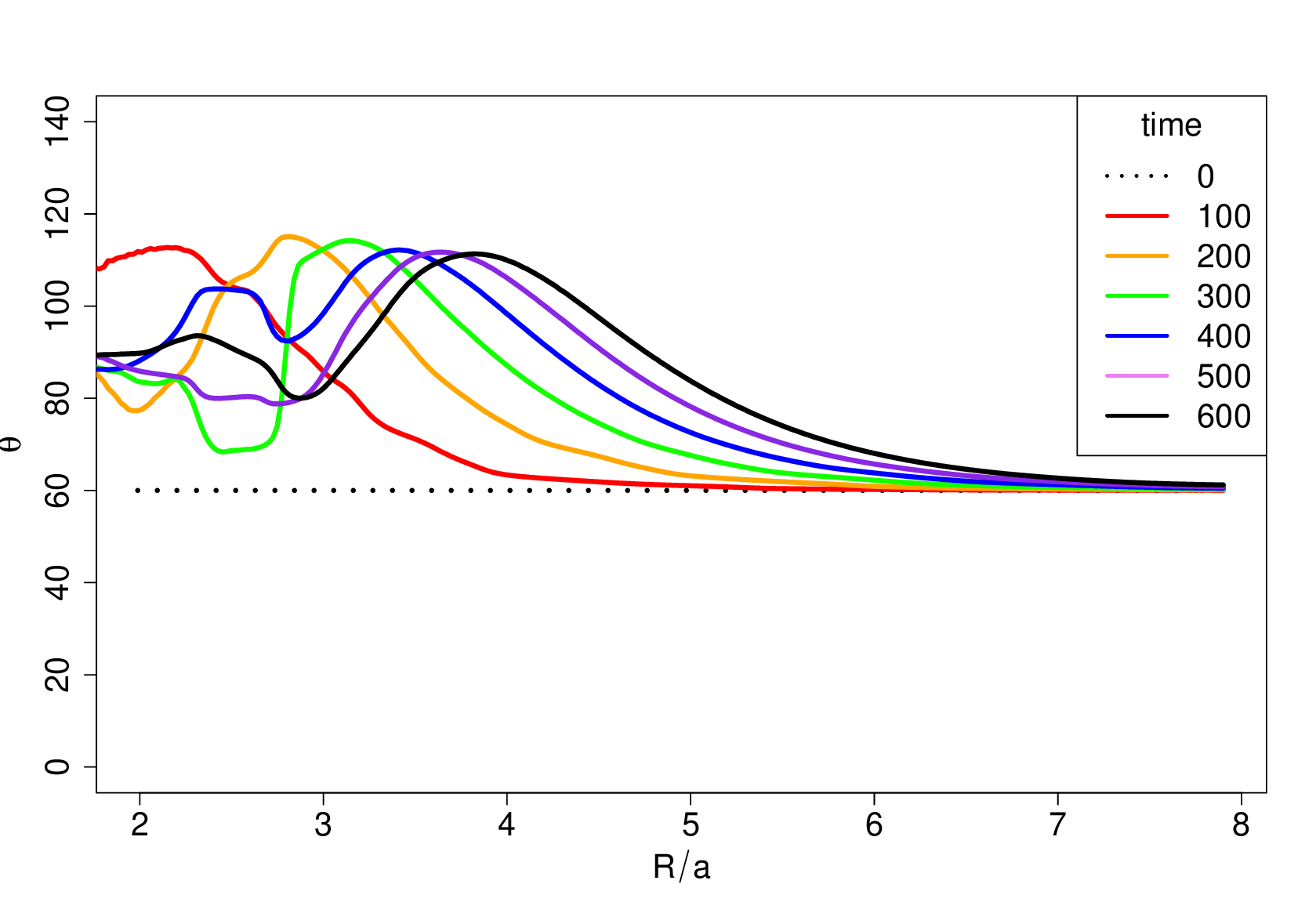}}
    \caption{
    \label{fig:tilt}Evolution of tilt profiles for $e=0$ (top) and $e=0.9$ (bottom) discs with initial tilt $\theta=60^\circ$ at $t=0$, 100, 200, 300, 400, 500, and 600 (see legend) in code units. }
  \end{center}
\end{figure}
%%%

%
% Fig. 5
%
\begin{figure*}
  \begin{center}
    \resizebox{42.0mm}{!}{\includegraphics{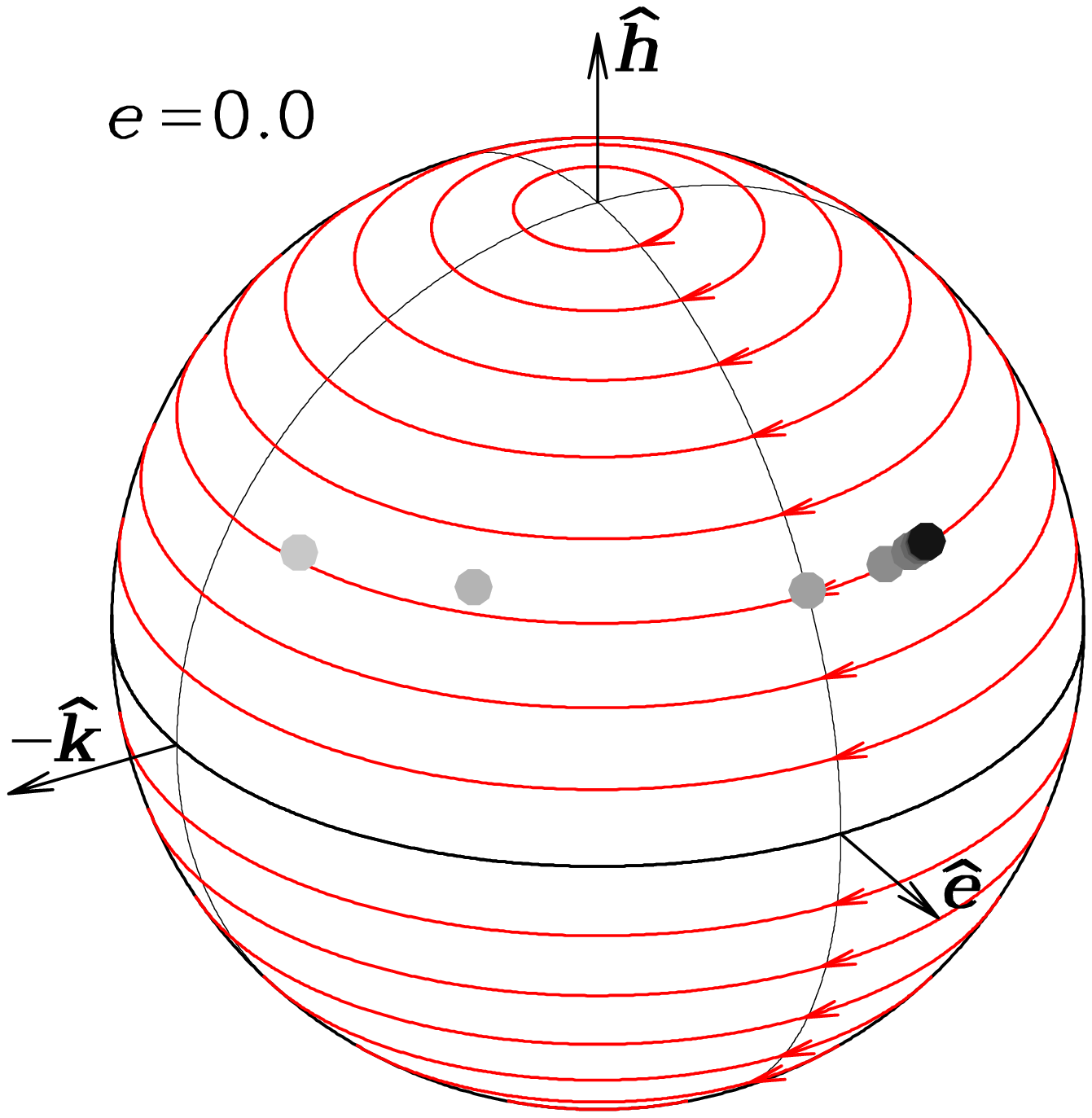}}\hfil
    \resizebox{42.0mm}{!}{\includegraphics{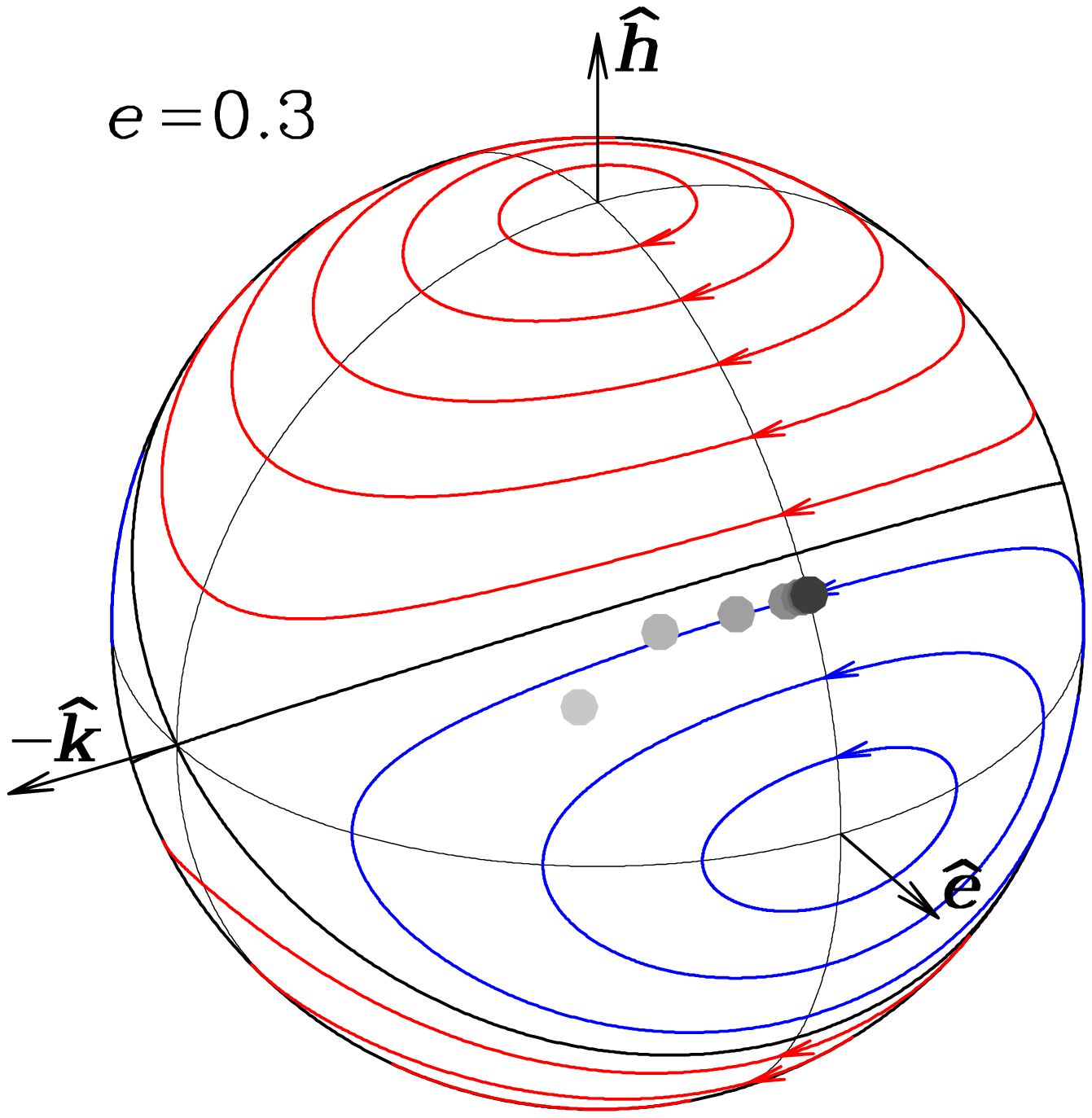}}\hfil
    \resizebox{42.0mm}{!}{\includegraphics{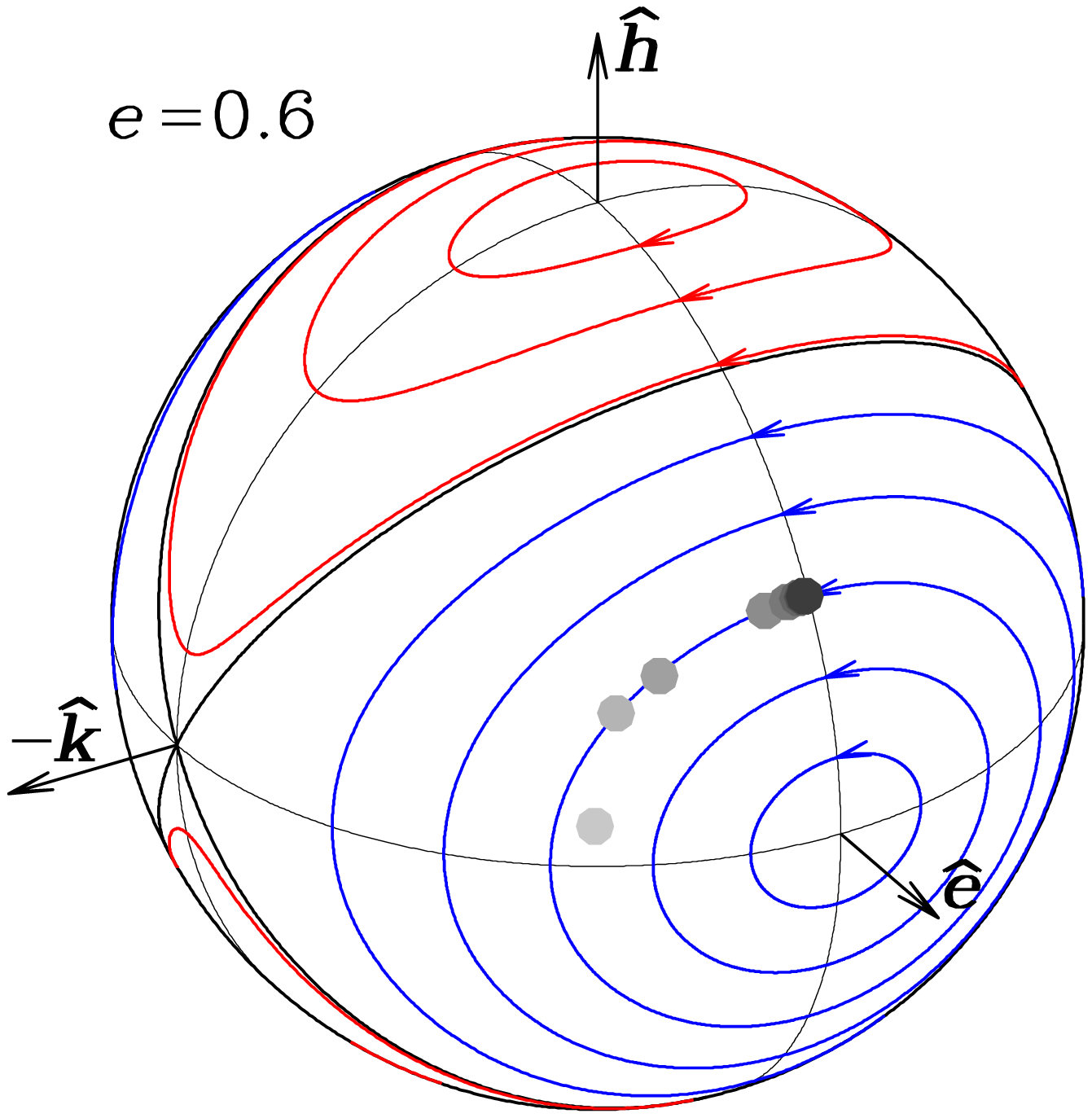}}\hfil
    \resizebox{42.0mm}{!}{\includegraphics{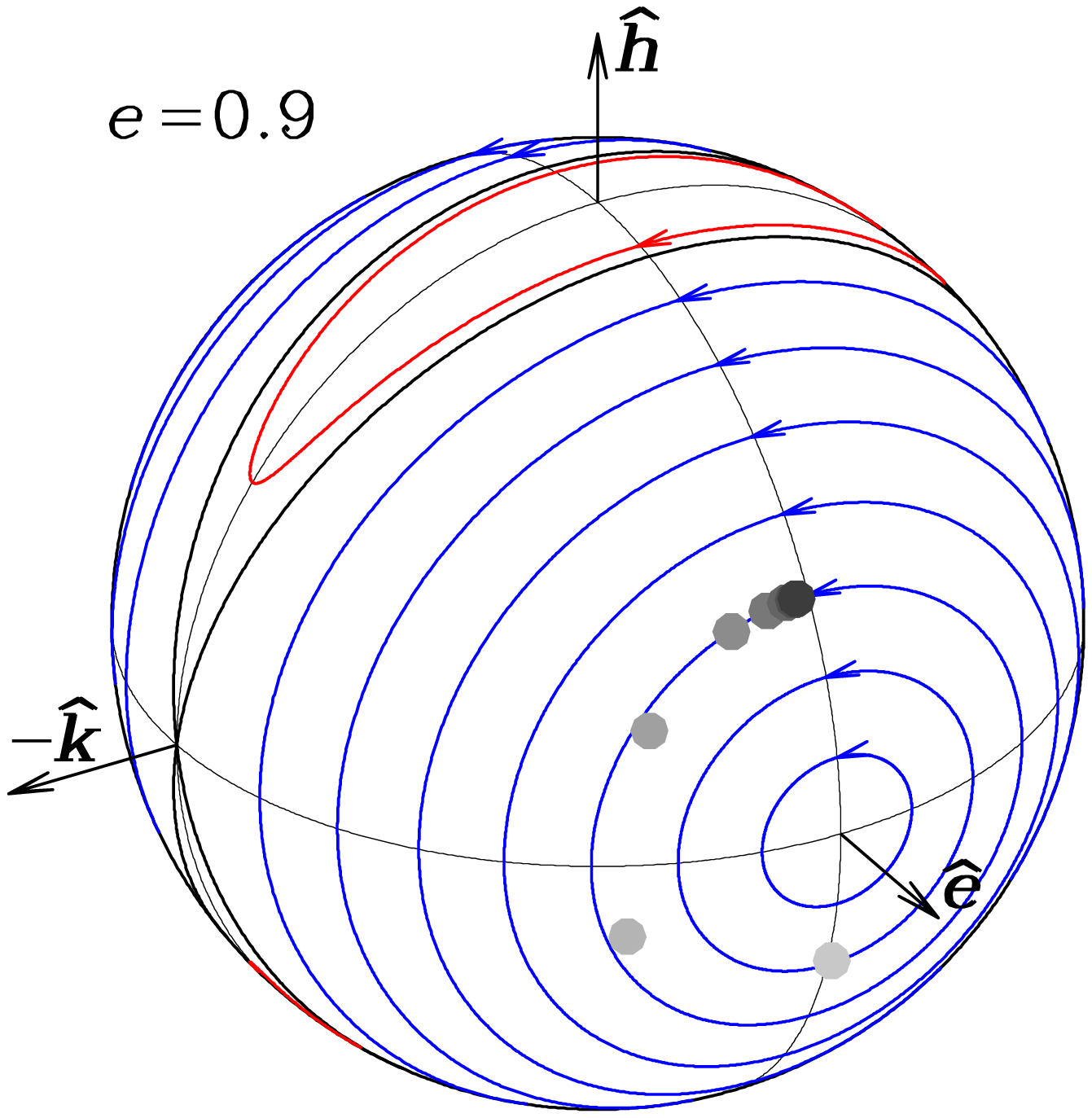}}
    \caption{
\label{fig:disc_prec} Projections of angular momenta of the radially binned disc in our simulations compared to the analytical precession paths shown in Fig.~\ref{fig:prec}. Solid circles represent seven radial bins of the disc ranging from $R=1$ (light gray) to $R=8$ (dark gray) at $t=100$ for simulations of different eccentricity (as indicated) but the same initial tilt $\theta=60^\circ$. Obviously, the inner disc precesses faster,
nicely following the theoretical precession paths. The innermost parts of the discs around eccentric binaries start to align with the stable polar orientation.}
  \end{center}
\end{figure*}
%%%

Fig.~\ref{fig:tilt} shows the time evolution of the tilt profiles $\theta(R)$ for two
simulations with initial $\theta=60^\circ$ but either $e=0$ (top) or $e=0.9$ (bottom).
For the circular binary case, the inner part of the disc eventually co-aligns with the binary ($\theta\to0^\circ)$, as expected. In contrast, for the highly eccentric binary we see the disc aligning in a polar configuration with respect to the binary angular momentum vector ($\theta\to90^\circ$). The variations in $\theta$ as function of both
time and radius are caused by the precession around $\H{e}$. 

This is more evident from Fig.~\ref{fig:disc_prec}, where we plot the orientation
$\H{l}$ of the angular momentum in annuli of the disc at $t=100$ for four simulations with different binary eccentricity but identical initial disc orientation at $\theta=60^\circ$. We see that the discs in our simulations closely follow the predicted precession paths especially in the outer parts of the disc. The inner parts of the disc, which have higher precession rates, dissipate faster and start to align with the $\H{h}$ or $\H{e}$ as expected. 

%%%%%%%%%%%%%%%%%%%%%%%%%%%%%%%%%%%%%%%%%%%%%%%%%%%%%%%%%%%%%%%%%%%%%%%%%%%%%%%%
\subsection{Violent ring interactions}
\label{sec:interaction}
Our simulations starting from discs misaligned to both the $\H{h}$ and $\H{e}$ show rather violent gas dynamics. The radially differential binary torque tears the disc and causes the formation of separate rings. These rings are mutually misaligned and start to interact with each other, presumably because they gained some eccentricity from interactions with the binary. The ring interactions cause partial cancellation of angular momentum and hence a significant increase in the accretion rate. This is identical to the picture reported by \cite{NixonEtal2013} for circumbinary discs around circular binaries. 

\begin{figure}
  \begin{center}
  \resizebox{41.7mm}{!}{\includegraphics[angle=0]{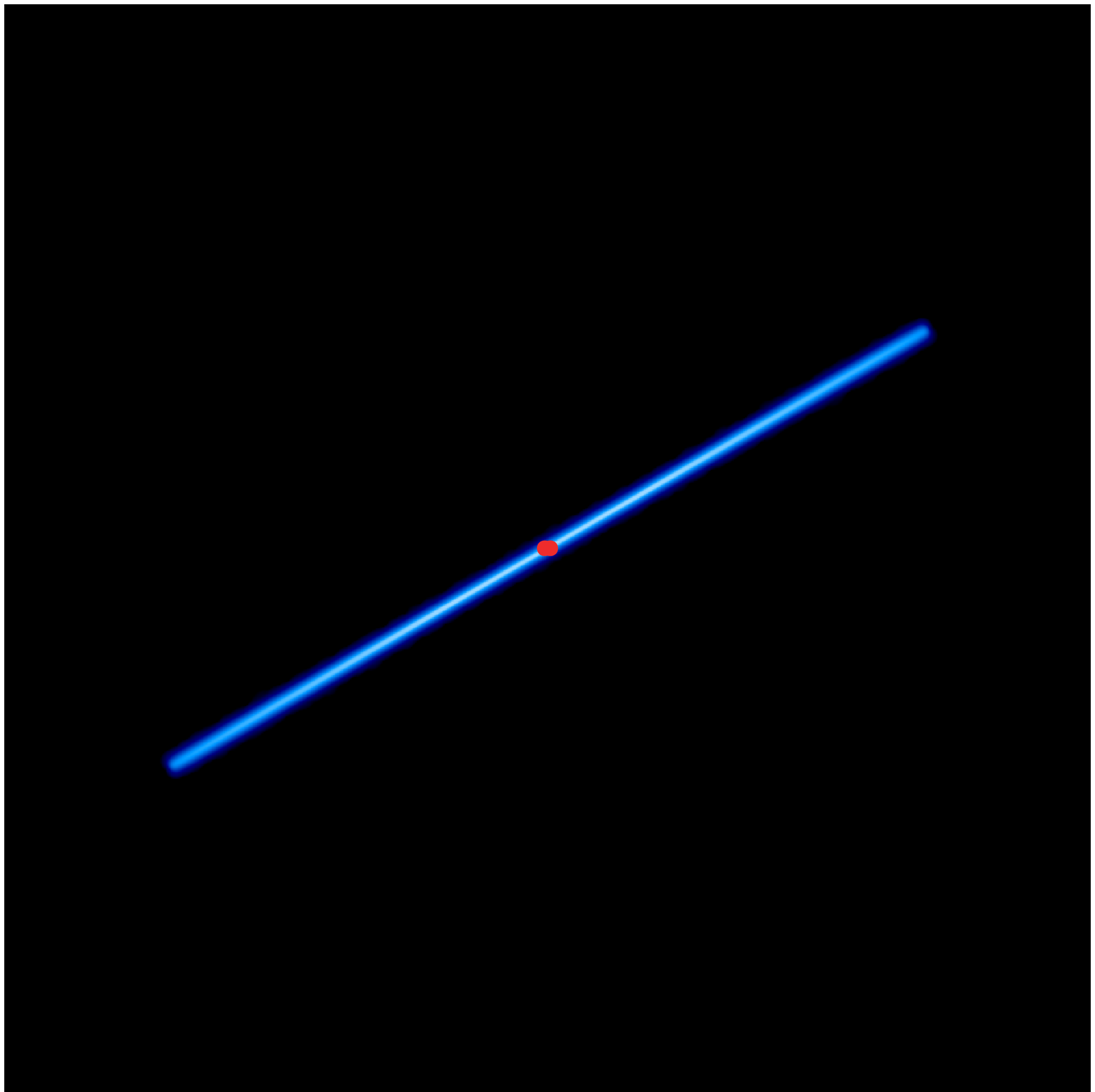}}
  \resizebox{41.7mm}{!}{\includegraphics[angle=0]{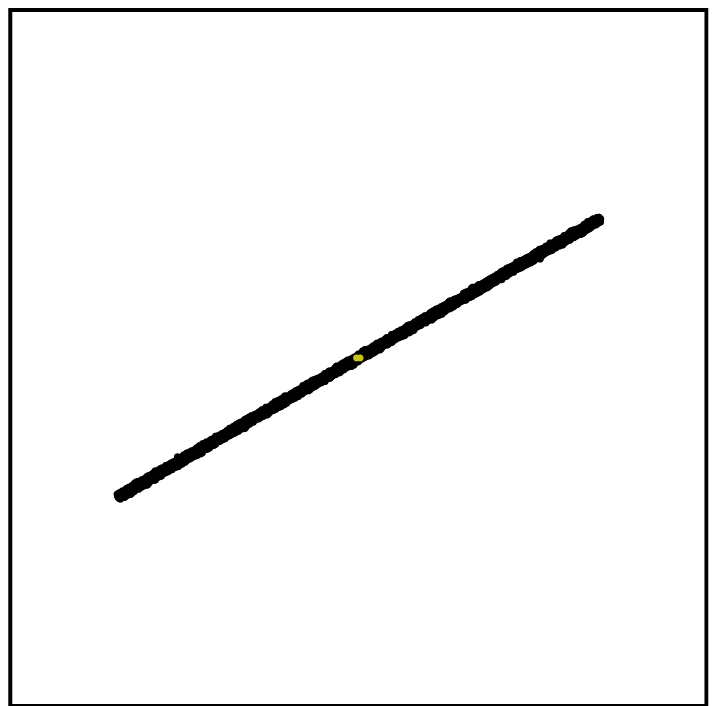}}
  \resizebox{41.7mm}{!}{\includegraphics[angle=0]{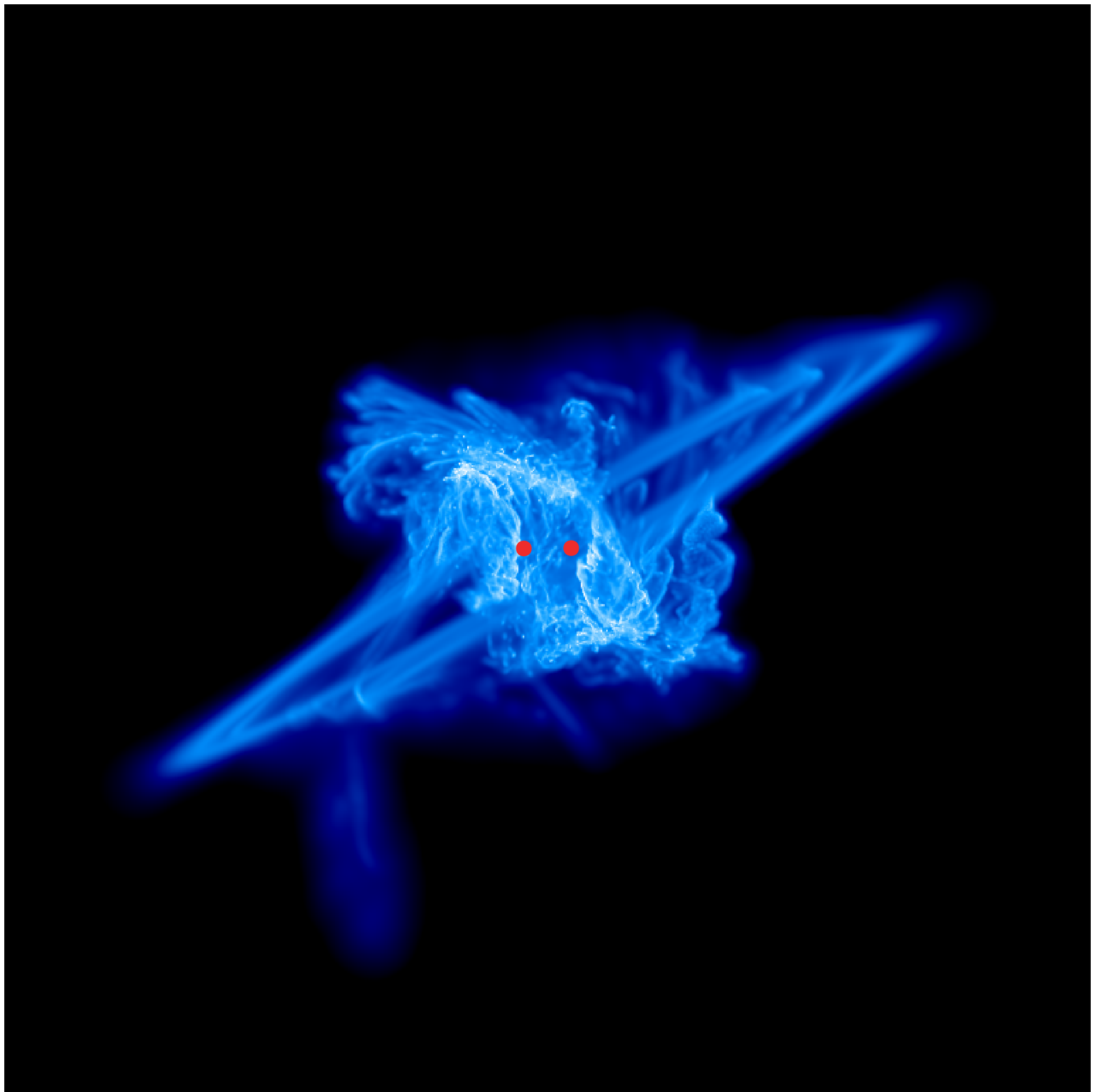}}
  \resizebox{41.7mm}{!}{\includegraphics[angle=0]{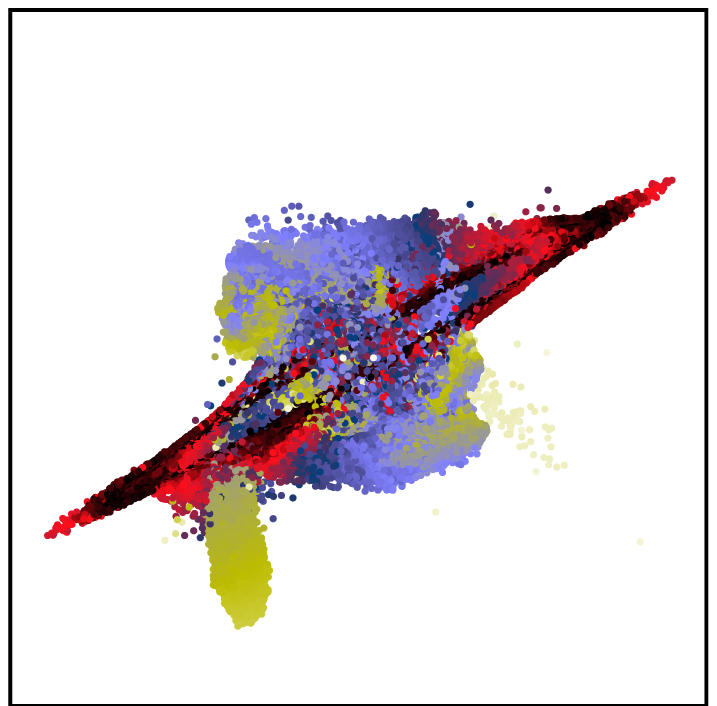}}
  \resizebox{41.7mm}{!}{\includegraphics[angle=0]{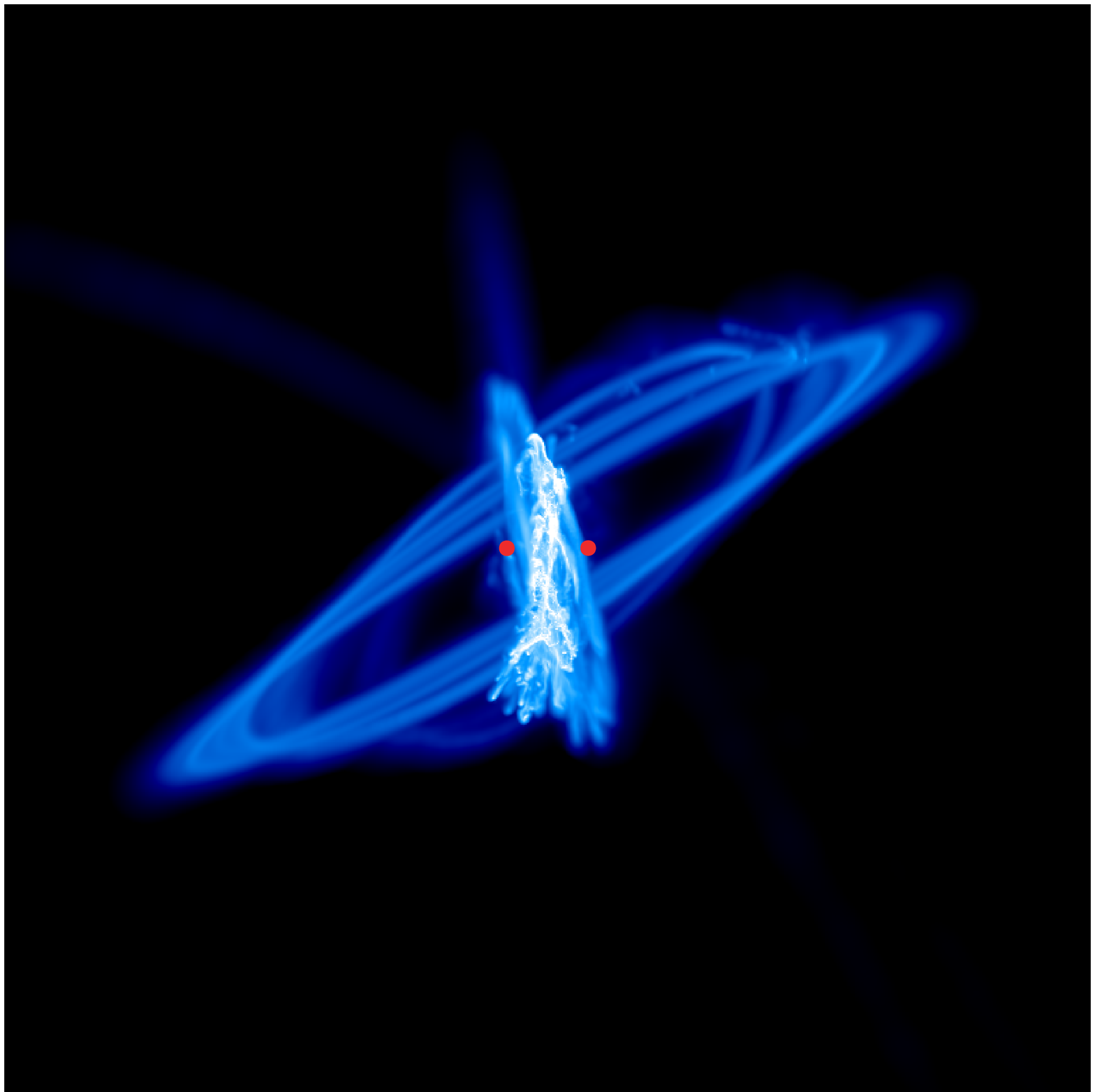}}
  \resizebox{41.7mm}{!}{\includegraphics[angle=0]{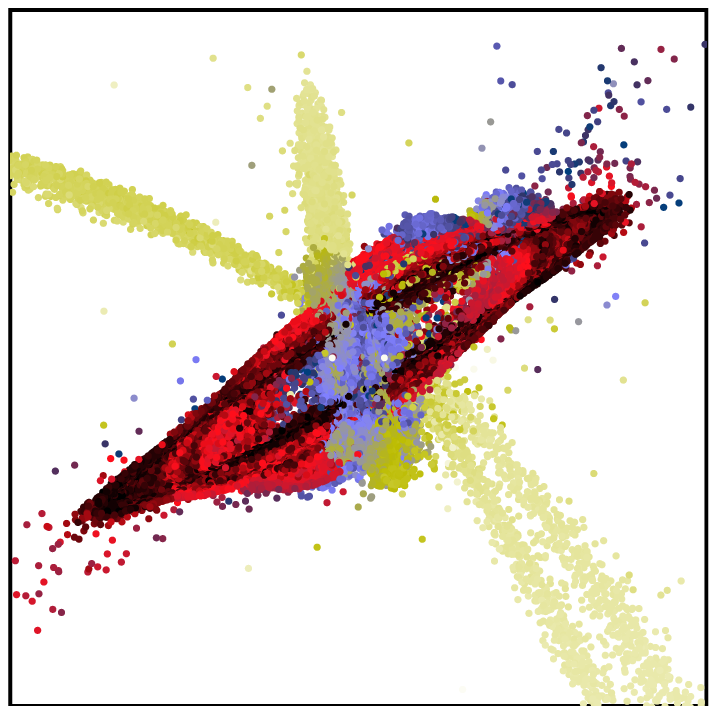}}
  \resizebox{41.7mm}{!}{\includegraphics[angle=0]{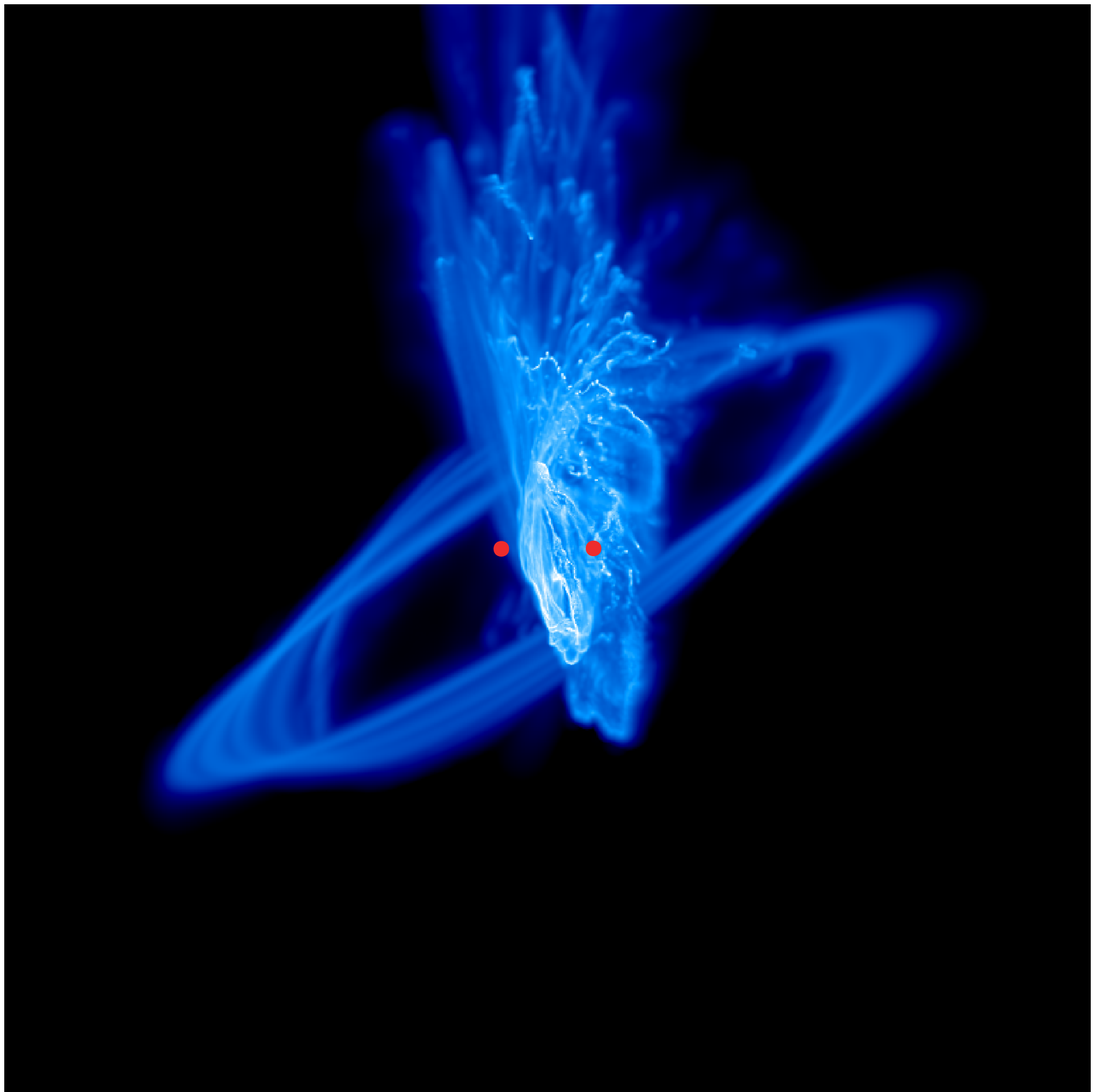}}
  \resizebox{41.7mm}{!}{\includegraphics[angle=0]{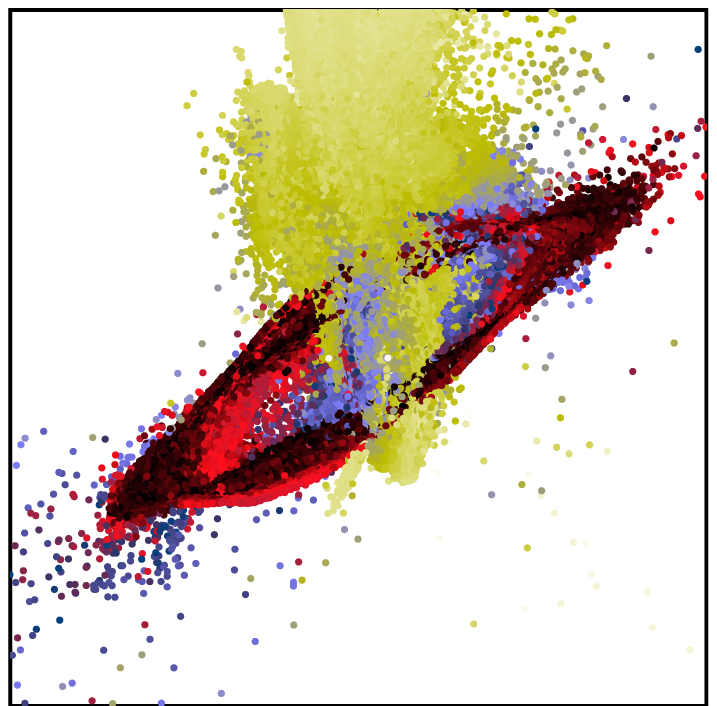}}
  \resizebox{41.7mm}{!}{\includegraphics[angle=0]{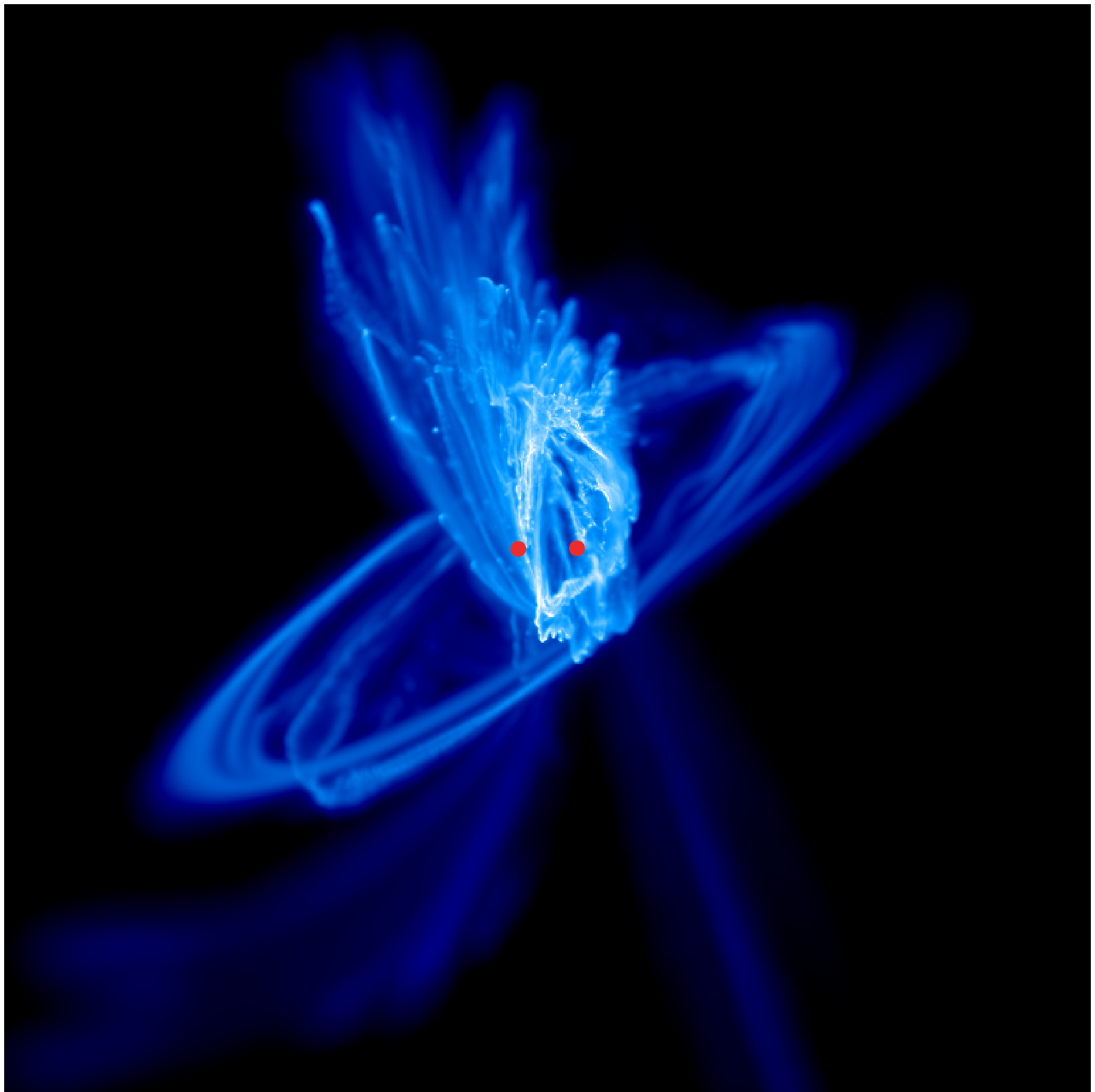}}
  \resizebox{41.7mm}{41.7mm}{\includegraphics[angle=0]{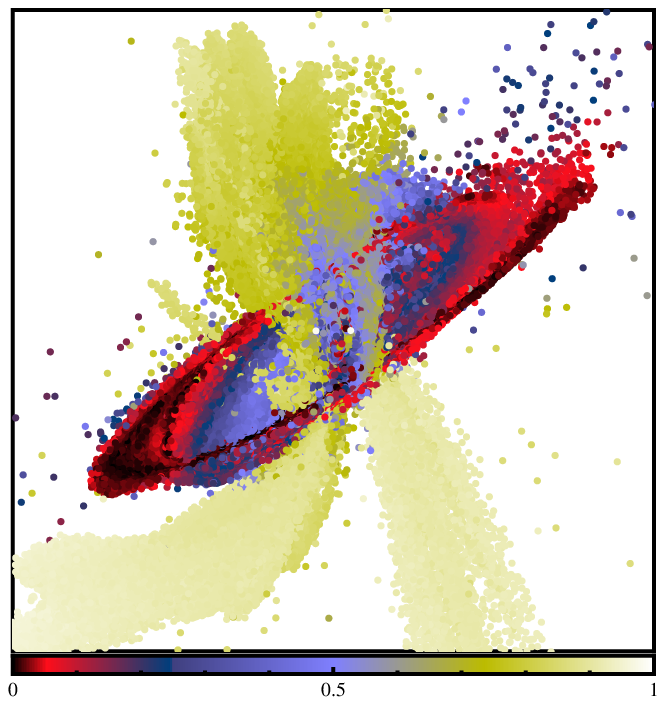}} 
  \caption{\label{fig:interaction} Density rendering (left panels) and particle plots coloured by eccentricity magnitude (right panels) of 5 snapshots of the $e=0.9$ $\theta=150^\circ$ run at times (from top to bottom) $t=0$, 200, 400, 600, and 800.  }
  \end{center}
\end{figure}

However, for very high eccentricities we find disc tearing to be much more violent and lead to a different evolution from that for circular and low-eccentricity binaries. There are two reasons for this difference: first the precession rate increases with eccentricity; second, the low-angular-momentum gas resulting from the interactions and falling onto the binary will align to polar orientation rather than a prograde or retrograde orientation as in the case of a near-circular binary. This allows this highly eccentric low-angular-momentum gas to come very close to the binary without suffering a lot of accretion. This non-circular gas in the central zone interacts with the outer disc further increasing its orbital eccentricity, throwing more gas to the centre, and promoting more interaction. This run away effect is shown in Fig.~\ref{fig:interaction} and can also be seen in the bottom left panel of Fig.~\ref{fig:render}. Eventually, this process sends an increasing amount of gas plunging onto the binary on almost radial orbits that can reach very close to the binary, avoiding significant accretion, and receiving energy kicks from one of the binary components in a manner very similar to the slingshot mechanism. Some of this gas will get ejected producing outward, almost radial, streams that can act as a possible observational signature of a highly eccentric SMBH binary.

Fig.~\ref{fig:interaction} shows density rendering (left panels) and particle plots coloured by eccentricity magnitude (right panels) of 5 snapshots for the simulation with $e=0.9$ and initially $\theta=150^\circ$ at times $t=0$, 200, 400, 600, and 800. We can see that the amount of chaotic gas resulting from the ring interactions keeps increasing during the simulation. The outward streams of gas resulting from the slingshot mechanism are very clear.
%
% Fig. 6
%
%%%

%
% Fig. 7
%
\begin{figure}
	\begin{center}
	\resizebox{81.7mm}{!}{\includegraphics{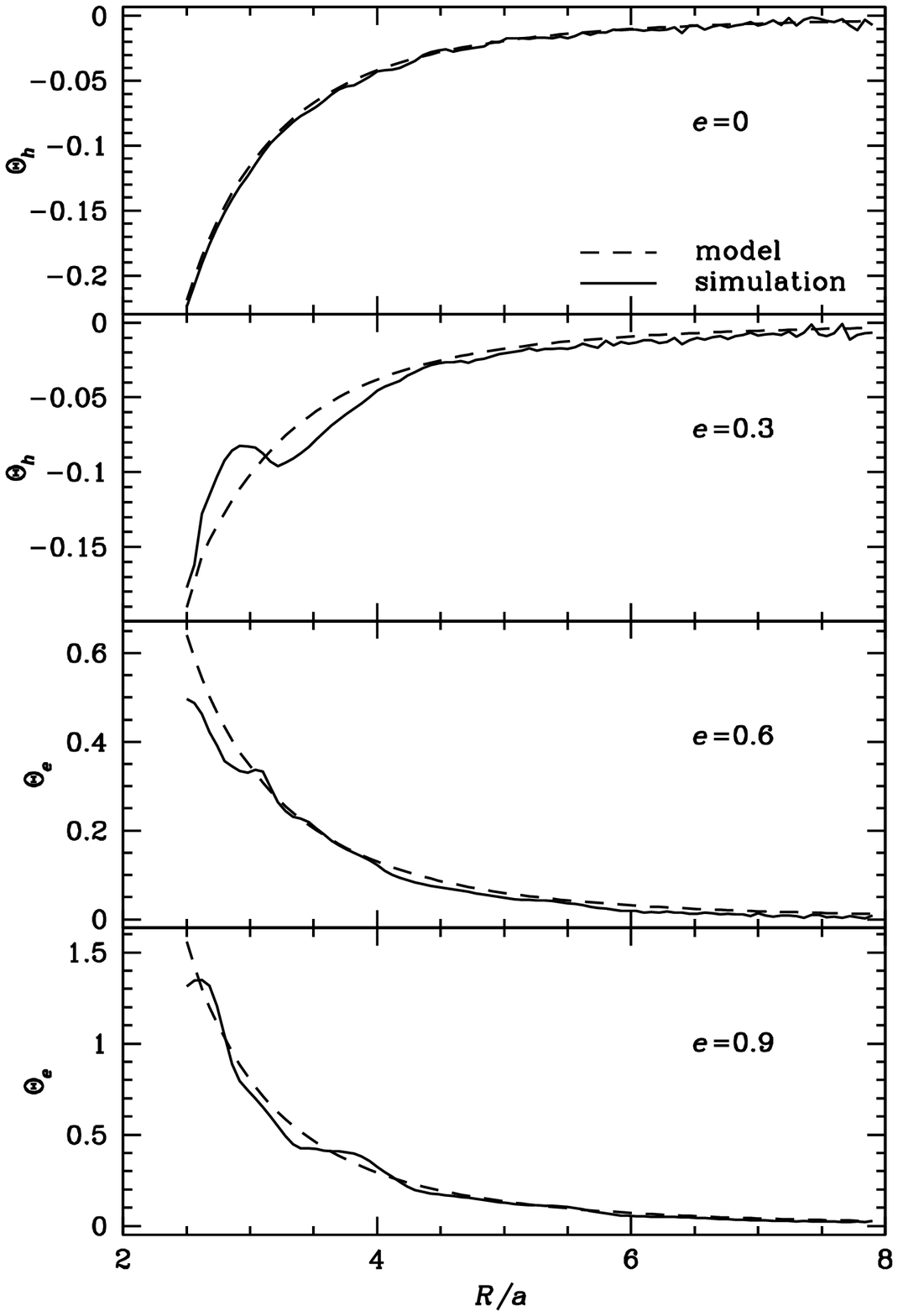}}
	\caption{\label{fig:precession_rate} Comparison between the disc precession rate
	measured from the simulation (solid) with initial $\theta=60^\circ$ and $e=0$, 0.3, 0.6, and 0.9 and our analytical model (dotted) of equation~(\ref{eq:Theta}). For each case we plot the dominant component of $\Theta$, i.e. $\Theta_h$ for $e=0$, 0.3 and $\Theta_e$ for $e=0.6$, 0.9.}
	\end{center}
\end{figure}
%%%

%%%%%%%%%%%%%%%%%%%%%%%%%%%%%%%%%%%%%%%%%%%%%%%%%%%%%%%%%%%%%%%%%%%%%%%%%%%%%%%%
\subsection{Precession rate}
In order to provide a quantitative comparison between the predictions of our analytical model in Section~\ref{sec:quadrupole} and the results obtained from the simulations, we plot in Fig.~\ref{fig:precession_rate} the analytical precession rate $\Theta$ derived from equation~(\ref{eq:Theta}) for a disc with an initial misalignment of $\theta=60^\circ$ around binaries with four different eccentricities along with the equivalent precession computed from the simulation and averaged over $10$ binary orbits starting from $t=50$. We find that the simulations agree quite well with the predicted precession rate at radii $\gtrsim 2.5 R/a$. For discs around eccentric binaries, we observe oscillations on binary orbital timescales and a good agreement is only found when the precession rate is averaged over a few binary orbits. We note that only a modest agreement is to be expected since our model ignores dissipative effects, contributions from higher than quadrupole order, and orbital resonances.

%%%%%%%%%%%%%%%%%%%%%%%%%%%%%%%%%%%%%%%%%%%%%%%%%%%%%%%%%%%%%%%%%%%%%%%%%%%%%%%%%%%%%%%%
\subsection{Non-zero initial disc twist angle}
So far, all the results shown here are for twist angles $\phi=0^\circ$, i.e.\ initially the disc line of nodes with the binary plane is the $\H{k}$ direction: $\H{l}$ is tilted towards $\H{e}$. In general, however, we should expect any disc orientation, i.e.\ non-zero $\phi$.

\begin{figure*}
\begin{minipage}{\textwidth}
\hspace{23.6 mm} $\theta=30^\circ$  \hspace{24.2 mm} $\theta=45^\circ$  \hspace{24.2 mm} $\theta=60^\circ$ \hspace{24.2 mm} $\theta=80^\circ$ \hspace{24.2 mm} $\theta=90^\circ$ 
\end{minipage}
\begin{minipage}[c]{0.06\textwidth}
\vspace{15 mm}$\phi=0^\circ$ \vspace{30 mm} \\ $\phi=45^\circ$ \vspace{30 mm} \\ $\phi=90^\circ$ \vspace{15 mm} \\ 
\end{minipage}%
\begin{minipage}{0.94\textwidth}
	\includegraphics[angle=0,width=\textwidth]{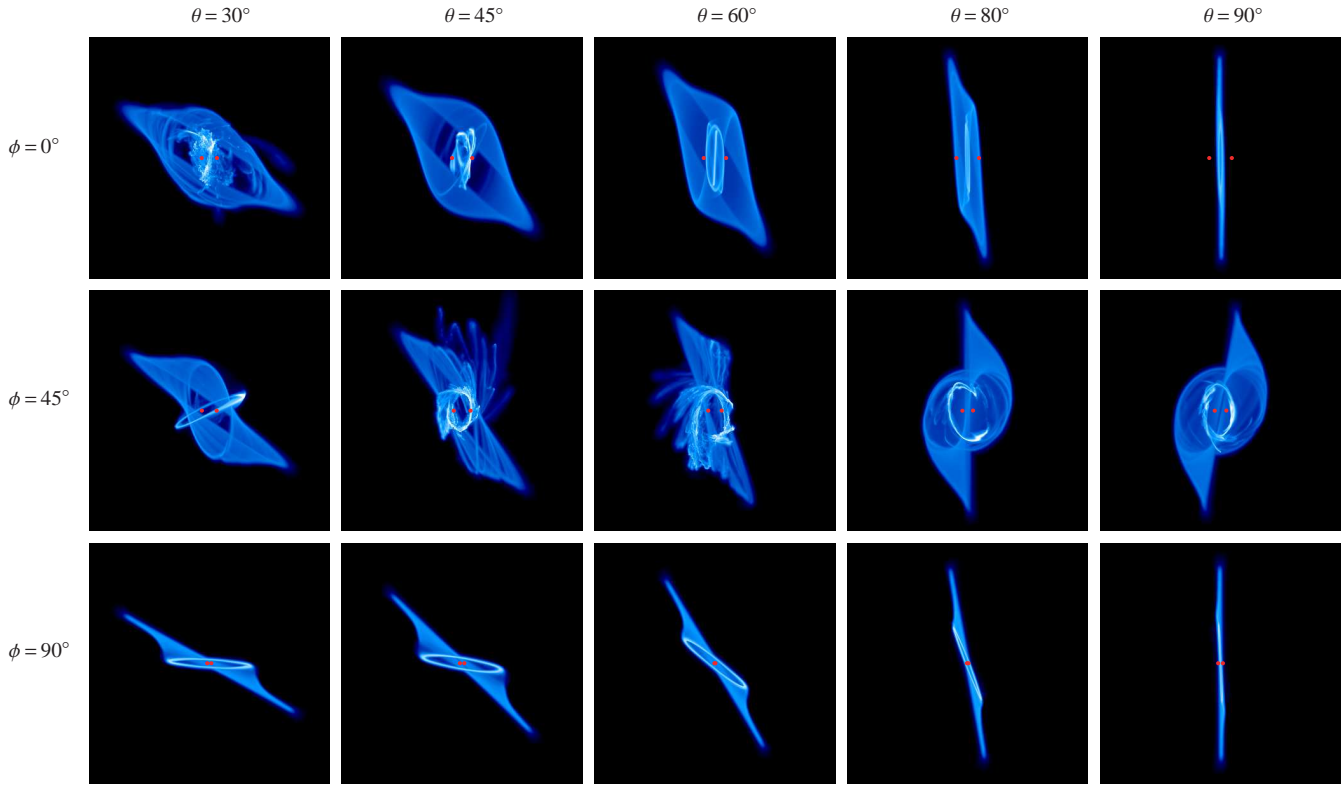}
\end{minipage}
	\caption{\label{fig:varying_phi}Density rendering of the $e=0.9$ simulations at $t=600$ ($\approx 95$ orbits of the binary) projected on the $x$-$z$ plane with different values for $\phi$ and $\theta$ as indicated in the figure.}
\end{figure*}

\begin{figure}
	\begin{center}
	\resizebox{55mm}{!}{\includegraphics{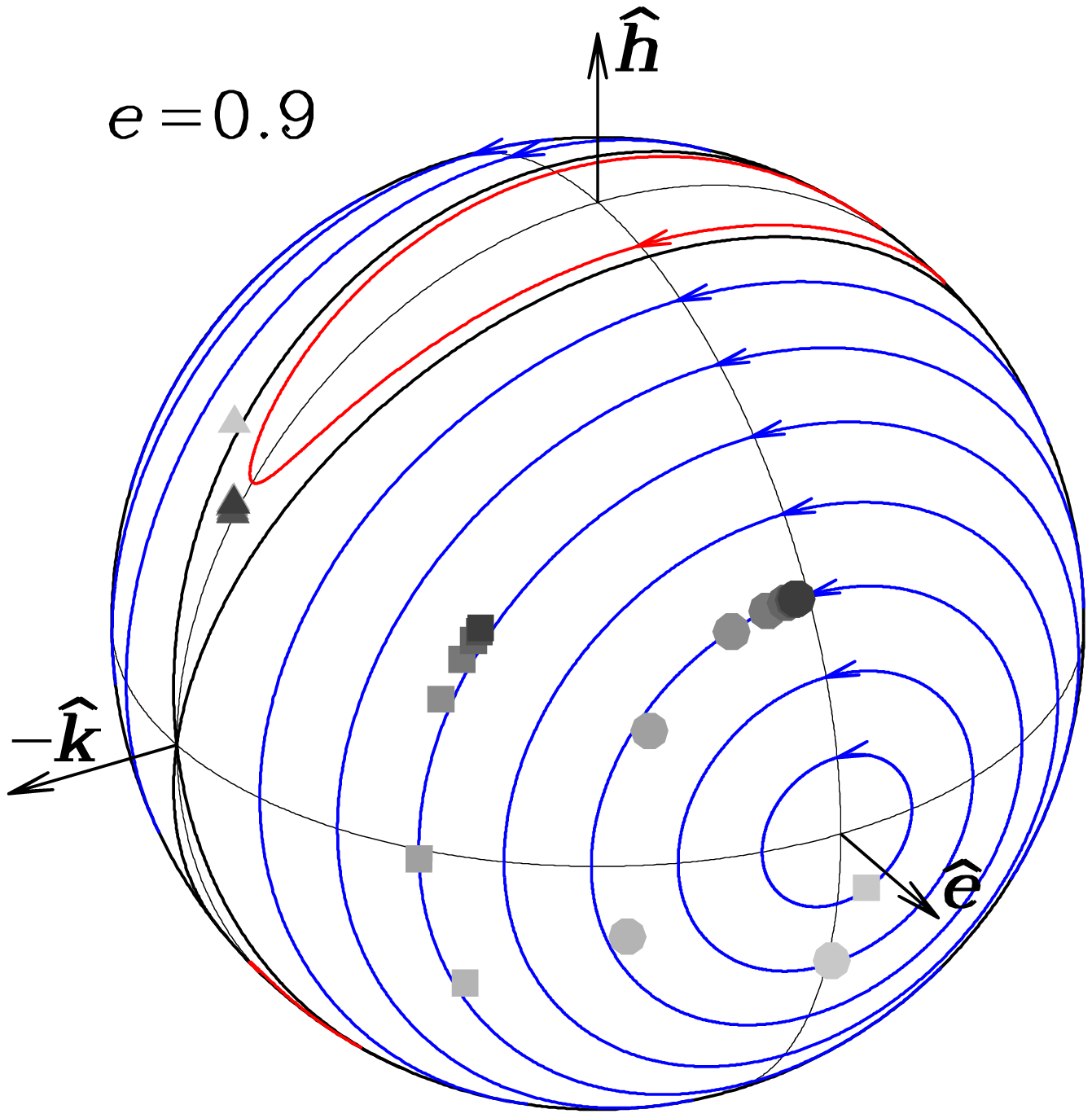}}
	\caption{\label{fig:precess_phi} Precession paths for $e=0.9$ and $\theta=60^\circ$ at $t=100$ with three different $\phi$ values: $\phi=0^\circ$ (circles), $\phi=45^\circ$ (rectangles), and $\phi=90^\circ$ (triangles).}
	\end{center}
\end{figure}

In Fig.~\ref{fig:varying_phi} we present snapshots for simulations with binary eccentricity $e=0.9$, initial twist angles $\phi=0^\circ$, 45$^\circ$, and 90$^\circ$, and initial tilt angles $\theta=30^\circ$, 45$^\circ$, 60$^\circ$, 80$^\circ$, and 90$^\circ$. For $\phi=90^\circ$, we only observe azimuthal precession, akin to the circular binary induced precession. This confirms our prediction since for that case $\H{l}$ and $\H{e}$ are always orthogonal, causing the first term in equation~(\ref{eq:Theta}) to vanish, and we are left with only azimuthal precession. For $\phi=45^\circ$, we find the same trend discussed earlier, i.e.\ experiencing either polar or azimuthal precession, or violent ring interaction. In Fig.~\ref{fig:precess_phi} we compare the precession paths of all three $\phi$ values for the case of $e=0.9$ and $\theta=60^\circ$ to the analytical contours. Similar to Fig.~\ref{fig:disc_prec}, we see the simulations closely follow the analytical contours apart from the innermost parts where disc disc breaking and aligment are dominant. This strongly suggests that our analysis above for $\phi=0^\circ$ carries over to the general case.

%
% Fig. 8
%
%%%

%
% Fig. 9
%
%%%

%%%%%%%%%%%%%%%%%%%%%%%%%%%%%%%%%%%%%%%%%%%%%%%%%%%%%%%%%%%%%%%%%%%%%%%%%%%%%%%%
\section{Implications for the final-parsec problem}
\label{sec:discuss}
The solution suggested by \cite{NixonEtal2011a} to the final parsec problem requires the binary to accrete negative angular momentum from a retrograde disc, which gradually increases the binary eccentricity until coalescence is achieved via energy losses to gravitational radiation at pericentre. \cite{NixonEtal2011b} showed that for a circular binary counter-alignment of randomly oriented accretion events can provide a continuous supply of the required retrograde discs. In particular, they showed that for cases where $J_{\mathrm{b}}>2J_{\mathrm{d}}$, all accretion events with initial misalignment of$\theta>90^\circ$ will result in a retrograde disc, i.e.\ roughly half of randomly oriented accretion events will counter-align with the binary as long as the binary dominates the angular momentum of the system.

Our results somewhat change this picture. As the binary eccentricity increases (due to retrograde accretion as in \citeauthor{NixonEtal2011a} or earlier stellar dynamical processes) disc counter- (and co-) alignment becomes ever less likely at the expense of polar alignment. The subsequent accretion of such polar discs merely rotates the angular momentum vector of the binary presumably hardly affecting the binary eccentricity. Thus simply retrograde gas accretion appears less viable a solution to the final parse problem.

There are however, still several ways gas can solve this problem. First, a single
massive retrograde accretion event may, in principle, supply enough negative angular momentum to complete the binary merger. However, for a massive disc self-gravity becomes important, likely causing clumping and star formation, which reduces the amount of gas that can be accreted. Moreover, a single massive retrograde accretion event may be not be sufficiently likely to explain the coalescence of all SMBH binaries (which form with
each major merger of massive galaxies).

A more intriguing possibility involves more violent gas dynamics. We showed that, in many cases, the disc does not smoothly align, instead the strong differential precession
(in particular for misaligned discs around eccentric binaries) leads to tearing of the disc into separate mutually misaligned rings. In the inner disc close to the binary, the gravity of the binary cause these rings to become eccentric such that they inevitably interact with each other and with the outer disc. These interactions cause further eccentricity growth on a dynamical time scale and eventually result in plunging gas infall. Some of this infalling gas will be accreted by either binary component.
This will change the binary angular momentum, but may not reduce its absolute value, depending on the orientation and in contrast to the situation with pre-dominantly retrograde accretion \citep{NixonEtal2013}.

If the infalling gas evades this fate, it will most likely get ejected from the binary via a slingshot interaction. This in turn reduces the binary separation in much the same way as the ejection of penetrating stars, thus exactly as required to solve the final parsec problem. Indeed, we find in our simulations which undergo violent gas dynamics not only significant gas accretion but also a shrinking of the binary orbit.

Clearly, this violent interaction and accretion processes are rather complex and chaotic and certainly not well resolved or adequately modelled in our simulations. Nonetheless,
what our simulations quite clearly show is that such violent gas-dynamical processes are inevitable if the gas is initially misaligned with the binary, in particular if the binary is eccentric. We leave a more detailed investigation into the binary orbital evolution in this chaotic environment for a future study.

%%%%%%%%%%%%%%%%%%%%%%%%%%%%%%%%%%%%%%%%%%%%%%%%%%%%%%%%%%%%%%%%%%%%%%%%%%%%%%%%
\section{Conclusions}
\label{sec:conclusion}
We have studied the interaction of an eccentric binary with a gaseous disc initially misaligned with the binary angular momentum. Such a configuration should occur naturally from the infall and subsequent circularisation of gas into the inner few parsec of a merger remnant still hosting a supermassive black hole (SMBH) binary \citep[e.g.][]{DunhillEtal2014}. The binary exerts a torque on the disc, resulting in disc precession and, due to viscous dissipation, in eventual alignment of the disc with the binary. In case of a circular binary, this alignment is always co-planar, resulting either in a pro- or retro-gradely rotating circumbinary disc \citep{NixonEtal2011b}.

We find that in the general case of an eccentric binary polar alignment also occurs, when disc angular momentum is aligned with the binary peri- or apo-apse direction. The binary torque on the disc can be quite accurately understood analytically assuming an orbit-averaged binary potential to quadrupole order (see Section~\ref{sec:quadrupole}). The fraction of initial disc orientations which give rise to polar alignment grows with binary eccentricity, reaching 0.5 at $e\approx0.4$. The precession paths (neglecting dissipation) are not circular, but elongated towards the intermediate axis of the orbit-averaged binary.

The prospect of accretion onto the binary from a polar instead of co-planar disc impedes the solution \citep[proposed by][]{NixonEtal2011a} to the final-parsec problem for coalescing a supermassive black-hole binary. In that picture consecutive randomly oriented accretion events lead to the formation of either pro- or retro-grade co-planar circum-binary discs. While accretion from the former is largely suppressed (by orbital resonances as discussed in the introduction), accretion from the latter reduces the binary angular momentum and drives it to larger eccentricities. However, at large eccentricities polar disc orientations dominate, when accretion (not resolved in our simulations) has presumably little effect on the binary orbit (since the accreted angular momentum is perpendicular to that of the binary). Thus, eccentricity growth via accretion is likely to be significantly reduced well before gravitational wave emission can take over as driver for coalescence.

However, in many of our simulations, in particular for larger binary eccentricity and stronger initial misalignment, the disc does not smoothly align, but is torn into separate mutually misaligned rings. This process was already reported by \cite{NixonEtal2013} for a circular binary and can be understood by the radially differential binary torque, which overcomes the adhesive effect of gas viscosity. The prominence of tearing with binary eccentricity and initial disc misalignment can be understood as consequence of the larger binary torque in these cases.

The subsequent evolution of these gas rings can be rather chaotic and is not quite adequately modelled in our simulations. However, some basic results appear to be robust. The innermost rings are sufficiently perturbed by the binary to acquire some orbital eccentricity. This in turn inevitably leads to interactions between the rings, resulting in partial cancellation of their angular momenta. This process is more prominent in
more eccentric binaries, because the stronger binary torque results in larger mutual misalignment between adjacent rings. The cancellation of angular momentum of the rings will increase their eccentricity, providing a positive feedback loop and hence a run-away process, eventually resulting in gas plunging onto the central binary.
This material may be accreted onto either hole, but when coming from a near-polar orientation, this will hardly help with the final-parsec problem, as explained above.

Alternatively, the infalling gas, which for a highly eccentric binary can come much closer to the binary whilst avoiding accretion, may get ejected from the binary via a gravitational slingshot interaction with one of its components. This also helps to solve the final-parsec problem, though this time by reducing its semi-major axis. This is similar to the stellar-dynamical process of shrinking the binary orbit via ejection of stars penetrating into the binary. The difference is that the total amount of stars in the `loss cone' (whose orbit carries them into inner parsec) is limited and cannot be easily re-filled, while gas being dissipative and collisional by nature may provide a better agent. This is particularly the case at the parsec scale where the SMBH dominates the dynamics and by its gravitational torques shepherds some gas into the loss cone.
%%%%%%%%%%%%%%%%%%%%%%%%%%%%%%%%%%%%%%%%%%%%%%%%%%%%%%%%%%%%%%%%%%%%%%%%%%%%%%%%
\section*{Acknowledgments}
We thank the referee for useful suggestions. Research in Theoretical Astrophysics at Leicester is supported by an STFC rolling grant. The calculations presented in this paper were performed using the ALICE High Performance Computing Facility at the University of Leicester. Some resources on ALICE form part of the DiRAC Facility jointly funded by STFC and the Large Facilities Capital Fund of BIS. CJN acknowledges support provided by NASA through the Einstein Fellowship Program, grant PF2-130098. We used {\sc splash} \citep{Price2007} visualization code for some of the figures in this paper.

%%%%%%%%%%%%%%%%%%%%%%%%%%%%%%%%%%%%%%%%%%%%%%%%%%%%%%%%%%%%%%%%%%%%%%%%%%%%%%%%
\bibliographystyle{mn2e} \bibliography{eccentric}

%%%%%%%%%%%%%%%%%%%%%%%%%%%%%%%%%%%%%%%%%%%%%%%%%%%%%%%%%%%%%%%%%%%%%%%%%%%%%%%%
\appendix
\section{Binary-Disc Quadrupole interaction}
\label{app:quadrupole}
Here, we give the details of the analysis leading to the results reported in Section~\ref{sec:quadrupole}. Our results, obtained via Newtonian dynamics, agree with the more general results of \cite{NaozEtal2013}, obtained via Hamiltonian perturbation theory, for a circular ring and ignoring octopole terms.

The three unit vectors $\H{h}$, $\H{e}$, and $\H{k}=\H{h}\cross\H{e}$ are conserved under the binary motion and are mutually orthogonal, such that
\begin{equation} \label{eq:identity}
	\hat{h}_i\hat{h}_{\!j} + 
	\hat{e}_i\hat{e}_{\!j} + 
	\hat{k}_i\hat{k}_{\!j} = \delta_{i\!j}.
\end{equation}
The binary components are at positions
\begin{equation}
	\B{x}_1 =  \frac{q}{1+q}\B{R},\quad
	\B{x}_2 = -\frac{1}{1+q}\B{R}
\end{equation}
with
\begin{equation}
	\B{R} = a(\cos\eta-e)\,\H{e} + a\sqrt{1-e^2}\sin\eta\,\H{k}.
\end{equation}
Here, $\eta$ is the eccentric anomaly, which is related to the mean anomaly $\ell$ via
\begin{equation}
	\ell = \eta - e \sin\eta,
\end{equation}
such that $\d\ell=(1-e\cos\eta)\d\eta$ and an orbit average becomes
$\langle\cdot\rangle=(2\pi)^{-1}\int_0^{2\pi}\cdot\,(1-e\cos\eta)\d\eta$. 
When orbit-averaging $R_iR_{\!j}$, the cross term between $\H{e}$ and $\H{k}$ averages to zero and
\begin{eqnarray}
	\A{R_iR_{\!j}}
	&=& \tfrac{1}{2}a^2\left[
	(1+4e^2)\hat{e}_i\hat{e}_j +
	(1- e^2)\hat{k}_i\hat{k}_j\right] \\
	&=&
	\tfrac{1}{2}a^2\left[
	5e^2\hat{e}_i\hat{e}_j +
	\left(1-e^2\right)\left(\delta_{i\!j}-\hat{h}_i\hat{h}_j\right)\right],
\end{eqnarray}
where the second form follows from eliminating $\hat{k}_i\hat{k}_{\!j}$ in favour of $\hat{h}_i\hat{h}_{\!j}$ with the help of the identity~(\ref{eq:identity}). From this result, we can work out the orbit-averaged trace-free specific quadrupole moment of the binary as
\begin{eqnarray}
	\mathsf{Q}_{i\!j}
	&=&
	M^{-1}\left[m_1\A{x_{1i}x_{1\!j}-\tfrac{1}{3}\B{x}_1^2\delta_{i\!j}}
		      + m_2\A{x_{2i}x_{2\!j}-\tfrac{1}{3}\B{x}_2^2\delta_{i\!j}}\right]
	\\ &=&
	\frac{q}{(1+q)^2} \left[
		\A{R_iR_{\!j}} - \tfrac{1}{3}\A{R_kR_{k}}
		\delta_{i\!j} \right]
	\\ 
	&=&
	\frac{a^2q}{(1+q)^2} \left[
	\left(\tfrac{1}{6}-e^2\right)\delta_{i\!j} +
	\tfrac{5}{2}e^2\,\hat{e}_i\hat{e}_{\!j} -
	\tfrac{1}{2}\left(1-e^2\right)\hat{h}_i\hat{h}_{\!j}\right].
\end{eqnarray}
We will also need the orbit average
\begin{equation} \label{eq:ave:RRV}
 	\A{R_iR_{\!j}\dot{R}_k} = \tfrac{1}{2} a^3\Omega e\sqrt{1-e^2}\left(
	\hat{e}_i\hat{k}_{\!j}\hat{e}_k +
	\hat{k}_i\hat{e}_{\!j}\hat{e}_k -
	2\hat{e}_i\hat{e}_{\!j}\hat{k}_k \right).
\end{equation}
with $\Omega=\sqrt{GM/a^3}$ the binary orbital frequency.

%%%%%%%%%%%%%%%%%%%%%%%%%%%%%%%%%%%%%%%%%%%%%%%%%%%%%%%%%%%%%%%%%%%%%%%%%%%%%%%%%%%%%
\subsection{Ring evolution}
\label{app:ringprec}
A ring particle at position $\B{r}$ experiences the orbit-averaged quadrupole potential of the binary
\begin{equation}
	\A{\Phi_{\mathrm{b}}}(\B{r}) = -\frac{3GM}{2r^5} \B{r}\cdot\S{Q}\cdot\B{r}.
\end{equation}
Averaging over the ring, we obtain 
\begin{equation}
	\A{r_ir_{\!j}}=\tfrac{1}{2}r^2 (\delta_{i\!j}-\hat{l}_i\hat{l}_{\!j}),
\end{equation}
such that the trace-free specific quadrupole moment of the ring is
\begin{equation}
	\mathsf{q}_{i\!j} = r^2 \left[\tfrac{1}{6}\delta_{i\!j} - \tfrac{1}{2}
	\hat{l}_i\hat{l}_{\!j}\right].
\end{equation}
The quadrupole interaction energy between binary and ring, averaged over the
the binary orbit and the ring, is then
\begin{eqnarray}
	\label{eq:E:app}
	\A{E_{\mathrm{br}}} &=& -\frac{3GMm}{2r^5} \mathrm{tr}(\S{\Theta})
\end{eqnarray}
with interaction tensor $\S{\Theta}\equiv\S{q}\cdot\S{Q}$, which has components
\begin{eqnarray}
	\label{eq:Theta:tensor}
	\mathsf{\Theta}_{i\!j} &=& \frac{a^2r^2q}{6(1+q)^2} \bigg[ 
	\left(\tfrac{1}{6}-e^2\right)\delta_{i\!j} + 
	\tfrac{5}{2}e^2\hat{e}_i\hat{e}_{\!j} -
	\tfrac{1}{2}(1-e^2)\hat{h}_i\hat{h}_{\!j} -
	\\[0.5ex] && \phantom{\frac{a^2r^2q}{6(1+q)^2}\bigg[}
	(\tfrac{1}{2}-3e^2)\hat{l}_i\hat{l}_{\!j} - 
	\tfrac{15}{2}e^2(\H{l}\cdot\H{e})\hat{l}_i\hat{e}_{\!j} + 
	\tfrac{3}{2}(1-e^2)(\H{l}\cdot\H{h})\hat{l}_i\hat{h}_{\!j}
	\bigg].\nonumber
\end{eqnarray}
Taking its trace in equation~(\ref{eq:E:app}) we obtain equation~(\ref{eq:E}).
The orbit-averaged torque on the ring also involves the interaction tensor. Using
index notation, we have
\begin{equation} \label{eq:l:dot}
	\dot{l}_i = - \varepsilon_{i\!jk} \A{r_j
	\frac{\partial\Phi_{\mathrm{b}}}{\partial r_k}}
	= \frac{3GM}{r^5} \varepsilon_{i\!jk}\, \mathsf{\Theta}_{\!jk}.
\end{equation}
Note that only the anti-symmetric part of $\S{\Theta}$ contributes to the torque.
Inserting equation~(\ref{eq:Theta:tensor}), we find that we can write this as
$\dot{\B{l}} = \B{\Theta} \cross \B{l}$ with the vector
\begin{equation}
	\B{\Theta} = \frac{3\omega q}{4(1+q)^2}\frac{a^2}{r^2}
	\left[
	5e^2(\H{l}\cdot\H{e})\,\H{e}
	-(1-e^2)(\H{l}\cdot\H{H})\,\H{h}
	\right],
\end{equation}
where $\omega=\sqrt{G(M+m)/r^3}$ is the ring angular frequency.
%%%%%%%%%%%%%%%%%%%%%%%%%%%%%%%%%%%%%%%%%%%%%%%%%%%%%%%%%%%%%%%%%%%%%%%%%%%%%%%%%%%%%%%
\subsection{Binary evolution}
The torque of the binary from the ring can be worked out analogously to that of the ring from the binary. The quadrupole potential due to the ring at the binary is 
\begin{equation}
	\Phi_{\mathrm{r}} = -\frac{3Gm}{2r^5} \B{x}\cdot\S{q}\cdot\B{x}.
\end{equation}
Adding the torque from each binary component and averaging over the orbit, we find
\begin{equation}
	\dot{h}_i = \frac{3Gm}{r^5} \varepsilon_{i\!jk}\, \mathsf{\Theta}_{k\!j}.
\end{equation}
In particular, the total angular momentum, $M\B{h}+m\B{l}$, is conserved at quadrupole order. Together with the precession of the ring, this implies that the evolution of $\B{h}$ is not simply a precession and that $h=|\B{h}|$ is in general not conserved. Instead, we find 
\begin{equation} \label{eq:h:dot:app}
	\dot{h} = -\frac{\omega^2 m}{\Omega M}
	\frac{15e^2h}{4\sqrt{1-e^2}}
	\,(\H{l}\cdot\H{e})\;(\H{l}\cdot\H{k}).
\end{equation}
Thus, $h$ remains unchanged only for a circular binary. $\dot{h}=0$ for $e=0$ or if $\H{l}$ is perpendicular to either $\B{e}$ or $\H{k}$. Otherwise, $h$ oscillates, since
$\H{l}\cdot\H{k}$ oscillates around zero under the ring precession.

The change of the eccentricity vector is
\begin{equation} \label{eq:e:dot:raw}
	\dot{\B{e}} = \frac{
	2\B{R}(\dot{\B{R}}\cdot\ddot{\B{R}}) -
	\dot{\B{R}}(\B{R}\cdot\ddot{\B{R}}) -
	\ddot{\B{R}}(\B{R}\cdot\dot{\B{R}})}{GM}
\end{equation}
with
\begin{equation} \label{eq:R:ddot}
	\ddot{\B{R}} = \frac{\partial\Phi_{\mathrm{r}}}{\partial \B{x}_2}-
	\frac{\partial\Phi_{\mathrm{r}}}{\partial \B{x}_1} =
	\frac{3Gm}{r^5} \B{R}\cdot\S{q}.
\end{equation}
Inserting~(\ref{eq:R:ddot}) into (\ref{eq:e:dot:raw}) and orbit averaging (using equation~\ref{eq:ave:RRV}), we find
\begin{eqnarray}
	\dot{\B{e}} \label{eq:e:dot:a}
	&=& 3\frac{\omega^2 m}{\Omega M}e\sqrt{1-e^2} \frac{2}{r^2} \left[
	\H{k}\,\H{e}\cdot\S{q}\cdot\H{e} -
	\H{e}\,\H{k}\cdot\S{q}\cdot\H{e} -
	\tfrac{1}{4}\H{k}\cdot\S{q} \right] \\	
	&=& 3\frac{\omega^2 m}{\Omega M}e\sqrt{1-e^2} \left[
	\left(\tfrac{1}{4} - (\H{l}\cdot\H{e})^2\right)\H{k}
	+(\H{l}\cdot\H{e})(\H{l}\cdot\H{k})\H{e}
	+\tfrac{1}{4}(\H{l}\cdot\H{k})\H{l}
	\right] \\
	&=&
	\frac{3}{4}\frac{\omega^2 m}{\Omega M}e\sqrt{1-e^2} \bigg\{
	\left[2-(\H{l}\cdot\H{h})^2-5(\H{l}\cdot\H{e})^2\right]\,
	\H{k}
	+\nonumber \\ &&
	\phantom{3\frac{\omega^2 m}{\Omega M}e\sqrt{1-e^2} \bigg\{}
	+(\H{l}\cdot\H{k})(\H{l}\cdot\H{h})\,\H{h}
	\label{eq:evec:dot}
	%\phantom{3\frac{\omega^2 m}{\Omega M}e\sqrt{1-e^2} \bigg\{}
	+5(\H{l}\cdot\H{e})(\H{l}\cdot\H{k})\,\H{e}
	\bigg\}.
%	\\
%	&=& 3\frac{\omega^2 m}{\Omega M}e\sqrt{1-e^2} \bigg[
%	\left(\tfrac{1}{4}+\tfrac{1}{4}(\H{l}\cdot\H{k})^2-(\H{l}\cdot\H{e})^2\right)\H{k}
%	+\tfrac{5}{4}(\H{l}\cdot\H{e})(\H{l}\cdot\H{k})\H{e}
%	+\nonumber \\ && \phantom{3\frac{\omega^2 m}{\Omega M}e\sqrt{1-e^2} \big[}
%	+\tfrac{1}{4}(\H{l}\cdot\H{k})(\H{l}\cdot\H{h})\H{h}
%	\bigg]\\
%	&=& 3\frac{\omega^2 m}{\Omega M}e\sqrt{1-e^2} \bigg\{
%	\tfrac{1}{4}\left[
%		\left(1+(\H{l}\cdot\H{k})^2-4(\H{l}\cdot\H{e})^2\right)\H{h} -
%		(\H{l}\cdot\H{k})(\H{l}\cdot\H{h})\H{k}
%	\right]\cross\H{e}
%	+\nonumber \\ && \phantom{3\frac{\omega^2 m}{\Omega M}e\sqrt{1-e^2} \bigg\{}
%	+\tfrac{5}{4}(\H{l}\cdot\H{e})(\H{l}\cdot\H{k})\H{e}
%	\bigg\}.
\end{eqnarray}
From this, we obtain
\begin{equation} \label{eq:e:dot:app}
	\dot{e} = \frac{15}{4} \frac{\omega^2 m}{\Omega M}e\sqrt{1-e^2}
	(\H{l}\cdot\H{e})(\H{l}\cdot\H{k})
\end{equation}
in agreement with equation~A34 of \cite{NaozEtal2013}, but also with 
equation~(\ref{eq:h:dot:app}) and $(1-e^2)\dot{h}=-he\dot{e}$ (from equation~\ref{eq:h:e}).

%%%%%%%%%%%%%%%%%%%%%%%%%%%%%%%%%%%%%%%%%%%%%%%%%%%%%%%%%%%%%%%%%%%%%%%%%%%%%%%%%%%%%%%
\subsection{Ring precession}
If the mass of the ring is negligible compared to that of the binary, we can approximate the binary orientation as fixed and the vectors $\hat{h}$, $\hat{e}$, and $\hat{k}$ as constants. In this case, the evolution of the ring orientation allows some further analytical treatment.

Since $\B{\Theta}$ is parallel to $\partial\A{E_{\mathrm{br}}}/\partial{\H{l}}$, precession is along lines of constant $\A{E_{\mathrm{br}}}$. This gives the equation
\begin{equation} \label{eq:klkl}
	C = (1-e^2)(\H{l}\cdot\B{h})^2 -5e^2(\H{l}\cdot\B{e})^2
\end{equation}
with constant $C$
for the precession paths. $C=0$ corresponds to the contour of $\A{E_{\mathrm{br}}}$
through the unstable orientations $\H{l}=\pm\H{k}$. Hence, this contour separates the regions of polar and azimuthal precession. The r.h.s.\ of equation~(\ref{eq:klkl}) can be written as $(\H{l}\cdot\B{u}_{1})\;(\H{l}\cdot\B{u}_{2})$ with
\begin{equation}
	\B{u}_{1,2}=\sqrt{1-e^2}\H{h}\,\pm\sqrt{5}\B{e}.
\end{equation}
Thus, the separatrices are great circles with poles $\B{u}_{1,2}$. The fraction of ring orientations undergoing polar precession is
\begin{equation}
	\frac{1}{\pi}\,\cos^{-1}(\hat{\B{u}}_1\cdot\hat{\B{u}}_2) = 
	\frac{1}{\pi}\,\cos^{-1}\frac{1-6e^2}{1+4e^2}.
\end{equation}
At small $e$, this grows linearly ($\propto\sqrt{20}e/\pi$) with eccentricity. Azimuthal and polar precession are equally likely for $\hat{\B{u}}_1\cdot\hat{\B{u}}_2=0$, which occurs at $e=6^{-1/2} \approx 0.408$.

If the constant $C$ in equation~(\ref{eq:klkl}) is positive (negative), we have
azimuthal (polar) precession. This equation for $\H{l}$ has the parametric solutions
for the precession paths (see also Fig.~\ref{fig:prec})
\begin{equation}
	\H{l}\cdot\H{e} = \sqrt{\frac{1-e^2-C}{1+4e^2}} \cos\zeta,
	\quad
	\H{l}\cdot\H{k} = \sqrt{\frac{1-e^2-C}{1-e^2}} \sin\zeta
\end{equation}
for $0<C\le1-e^2$ (azimuthal precession), and
\begin{equation}
	\H{l}\cdot\H{h} = \sqrt{\frac{5e^2+C}{1+4e^2}} \cos\zeta,
	\quad
	\H{l}\cdot\H{k} = \sqrt{\frac{5e^2+C}{5e^2}} \sin\zeta
\end{equation}
for $-5e^2\le C<0$ (polar precession). In either case, the third component of $\H{l}$ follows from the normalisation condition $|\H{l}|=1$.

The instantaneous precession rate $|\d\H{l}/\d t|=|\B{\Theta}\cross\H{l}|$ varies along the precession paths. It is minimal at the largest value of $|\H{l}\cdot\H{k}|$ along the precession path and drops to zero for $\H{l}=\H{k}$ (the unstable orientations).
The instantaneous precession rate becomes maximal at $\H{l}\cdot\H{k}=0$ (at  zero twist $\phi$), when
\begin{equation}
	|\d\H{l}/\d t|_{\max}=\frac{3\omega q}{4(1+q)^2}\frac{a^2}{r^2}
	(1+4e^2)\sin\theta\cos\theta.
\end{equation}
Thus, the maximum precession rate is much larger for highly eccentric than for circular
binaries. The largest variation of the precession rate occurs for precession paths close to the separatrices.

\label{lastpage}
\end{document}